\documentclass[10pt]{article}
\usepackage{amsmath}
\usepackage{graphicx}
\usepackage{url}
\usepackage{multicol}
\usepackage{cite} 
\usepackage{afterpage} 
\usepackage{geometry} 
\usepackage{float}

\usepackage{epsfig}
\usepackage{color}
\usepackage[dvipsnames]{xcolor}
\graphicspath{{FIGURES/}} 

\usepackage[aboveskip=1pt,labelfont=bf,labelsep=colon,justification=justified,singlelinecheck=off]{caption}
\renewcommand{\figurename}{Fig}

\usepackage{cite}
\usepackage{hyperref}
\makeatletter
\renewcommand{\@biblabel}[1]{\quad#1.}
\makeatother

\usepackage{longtable}
\usepackage{array}
\newcolumntype{L}[1]{>{\raggedright\let\newline\\\arraybackslash\hspace{0pt}}m{#1}}
\newcolumntype{P}[1]{>{\raggedright\arraybackslash}p{#1}}

\title{\LARGE Self-organization and time-stability of social hierarchies}
\date{}
\author{Joseph Hickey\footnote{joseph.hickey@ucalgary.ca} $\/$ and J{\"{o}}rn Davidsen \\\\{{\normalsize Complexity Science Group}}\\{\normalsize Department of Physics and Astronomy}\\{\normalsize University of Calgary}}

\begin{document}
\pagenumbering{arabic}

\maketitle

\section*{Abstract}

The formation and stability of social hierarchies is a question of general relevance. Here, we propose a simple generalized theoretical model for establishing social hierarchy via pair-wise interactions between individuals and investigate its stability. In each interaction or fight, the probability of ``winning'' depends solely on the relative societal status of the participants, and the winner has a gain of status whereas there is an equal loss to the loser. The interactions are characterized by two parameters. The first parameter represents how much can be lost, and the second parameter represents the degree to which even a small difference of status can guarantee a win for the higher-status individual. Depending on the parameters, the resulting status distributions reach either a continuous unimodal form or lead to a totalitarian end state with one high-status individual and all other individuals having status approaching zero. However, we find that in the latter case long-lived intermediary distributions often exist, which can give the illusion of a stable society. As we show, our model allows us to make predictions consistent with animal interaction data and their evolution over a number of years. Moreover, by implementing a simple, but realistic rule that restricts interactions to sufficiently similar-status individuals, the stable or long-lived distributions acquire high-status structure corresponding to a distinct high-status class. Using household income as a proxy for societal status in human societies, we find agreement over their entire range from the low-to-middle-status parts to the characteristic high-status ``tail''. We discuss how the model provides a conceptual framework for understanding the origin of social hierarchy and the factors which lead to the preservation or deterioration of the societal structure.

\newpage

\section{Introduction} \label{section:intro} 

\paragraph{} Animals, including humans, form social hierarchies \cite{Sapolsky2005, Chiao2010, Chase2011, Marmot2014, Boyce2004}. How these hierarchies form and what makes them remain stable over time is a central question across many different fields. In the humanities, social and political theorists have studied the origin of class structures and the conditions under which these structures are preserved or change \cite{Hobbes1998, Lagasse2017, Shayo2009, Kunst2017}. Archaeologists and other researchers from diverse fields study the factors that lead to the collapse of civilizations \cite{Butzer2012, Diamond2010}. Anthropological research has focused on the roles of norms, sanctions, and cooperative behaviour in creating and maintaining hierarchy \cite{Rand2013, Fehr2004, Jensen2010}. In the biological sciences, researchers have questioned whether hierarchy emerges primarily from differences in intrinsic qualities of individuals (e.g. physical strength, intelligence, or aggressive tendency) or as a self-organizing process in which a hierarchy arises as a result of many interactions between the members of the society \cite{Chase2002, Franz2015, Rutte2006, Lindquist2009}.

\paragraph{} From a high-level perspective, a fundamental question arises: Can a stable or long-lived hierarchical structure occur entirely by self-organization, based solely on inter-individual interactions, modeled as independent pair-wise ``fights"? And, if so, what are the typical structures of hierarchy, and what are the characteristic times of formation and evolution of the said structures?

\paragraph{} ``Winner-loser" models are a class of mathematical models that have been used to study the self-organization of social hierarchy in biology \cite{Hsu2006, Lindquist2009, Bonabeau1995, Bonabeau1999, Odagaki2006, Odagaki2007, Ben-Naim2005, Ben-Naim2006, Albers2001, Dugatkin1997, Hogeweg1983, Hemelrijk1999} and economics \cite{Yakovenko2009, Ispolatov1998, Hayes2002, Boghosian2017, Chakraborti2011}. In these models, individuals are characterized by a property, such as ``strength", ``resource holding potential", or ``wealth",  that determines the individual's position in society (in the following, we use ``strength" as a generic term for this property). Pairs of individuals come into contact and engage in an interaction (or ``fight"). The fight has a winner and a loser, where the winner experiences a gain in strength, and the loser loses strength. The models have two basic rules: one that determines who wins in a given fight, and another that determines the amount of strength gained or lost in a fight. The distribution of strength, which changes as individuals interact with each other, represents the societal structure resulting from the model. While stable societal structures have been analyzed in previous studies of winner-loser models, the time evolution and intermediary, potentially long-lived societal structures have been mostly neglected. Here, we aim to close this crucial knowledge gap.

\paragraph{} To do so, we construct a generalized winner-loser model in which we intend the strength property to represent societal status. The amount of status gained by the winner and lost by the loser of each fight is proportional to the pre-fight status of the losing individual. We define a probability for winning that is determined by the relative statuses of the two competitors, modulated by a parameter spanning a continuous range of degree of authoritarianism from redistributive (lower-status opponent always wins) to totalitarian (higher-status opponent always wins). The latter modulation for winning contains previous models as special cases at specific values of the authoritarianism parameter, and allows a more general description of the dynamics. Over a large range of parameters and excluding these special cases, we find the emergence of long-lived intermediary societal structures (distributions of societal status) for the first time. Establishing the existence of these long-lived structures --- which can give the illusion of a stable society --- and the relationship between the characteristic time of their evolution and the model parameters is one of the main contributions of our study.

\paragraph{} To demonstrate the relevance of our generalized model and the long-lived structures, we analyze real-world data. Specifically, we compare data from observational studies on wins and losses in animal interactions with the results from simulations of our model, and we compare the distributions of societal status produced by the model with real-world social hierarchies. To make the latter comparison, we use proxies for societal status in large social groups. In both cases, the real-world data are consistent with our model. Specifically, in our model, long-lived intermediary societal structures (distributions of societal status) arise independent of whether any pair of individuals are equally likely to interact or not. In the latter case, however, status distributions with more complex shapes consistent with the household income proxy emerge. We are able to fit the simulated status distributions to USA household income data with good agreement. To our knowledge, this is the first model that produces the two-part structure of the proxy distribution by self-organization based solely on interacting individuals. 

\paragraph{} The model is presented in section \ref{section:the_model} and an extended version of it in which similar-status individuals interact more frequently than individuals with large differences in status is presented in section \ref{section:extended_model}. Details about the shapes of the status distributions and their evolution in time are presented in section \ref{section:struct_and_time_evol}. Comparison of model results to data from real societies is contained in section \ref{section:realworld}, where we consider data on agonistic interactions in non-human animals in section \ref{section:agonistic} and proxies for societal status in large social groups (social insects and humans) in section \ref{section:proxies}. The article concludes with a summary of results and some comments regarding future research directions.

\section{Definition of the model} \label{section:the_model}

\paragraph{} Winner-loser models have been constructed using many variations of the rule determining who wins the fight and the rule determining the amount of strength gained or lost in a fight, where the particular formulation chosen for each rule depends on the system under study. 

\paragraph{} In the rule determining the amount of strength gained and lost in a fight, two formulations have been applied previously. In one version (``additive'' rule), the effect of fighting on an individual's strength accumulates additively, for example, by the addition or subtraction of a fixed increment of status \cite{Bonabeau1995, Bonabeau1999, Odagaki2006, Odagaki2007, Ben-Naim2005, Ben-Naim2006}. In an additive rule, the amount of strength gained or lost in a fight does not depend on the current value of either individual's strength. This means that the amount of strength won or lost in a fight is always the same, regardless of the strength of one's opponent.

\paragraph{} The other version of this model rule is a ``multiplicative'' one. Here, the amount of strength gained or lost is proportional to the strength of one of the individuals involved in the fight, such that effect of fighting accumulates multiplicatively \cite{Dugatkin1997, Ispolatov1998, Hayes2002, Boghosian2017}. Defeating a strong opponent produces a large increase in strength, whereas defeating a weak opponent produces a small increase in strength. It is clear from animal behaviour studies that wins against high ranking individuals increase the rank of an individual more than wins against low ranking individuals. In this case, a multiplicative rule is therefore more realistic than an additive rule, in which it is no more advantageous for an individual to defeat a strong rather than a weak rival. Moreover, whether an additive or multiplicative rule is used leads to substantially different distributions of strength \cite{Hsu2006}. For example, in many models with additive rules, strength becomes distributed such that individuals of adjacent ranks are separated by the same amount of strength. In multiplicative models, on the other hand, highly skewed distributions can result, and such multiplicative processes have been proposed as a common underlying cause of observed inequalities in natural and social systems \cite{Scheffer2017, Sutton1997}. 

\paragraph{} Here, we implement a formulation of the multiplicative rule in which the amount of strength won or lost is proportional to the pre-fight status of the losing individual (``loser scheme''). In another formulation that has been used in several econophysics models \cite{Hayes2002, Chakraborti2011, Boghosian2015, Boghosian2016, Boghosian2016b, Boghosian2017}, the amount of strength won or lost is proportional to the pre-fight status of the weaker individual, regardless of who wins or loses (``poorer scheme''). The loser scheme formulation is more realistic in the context of dominance hierarchies, because upset victories, in which the lower-strength individual in the pair wins, produce large rewards for the winner and large penalties for the loser. For example, in primate dominance hierarchies, only a small number of repeated defeats of a higher-strength individual by the same lower-strength individual are required for their rankings to be reversed \cite{Gesquiere2011}. This scenario is captured by the loser scheme but not by the poorer scheme.

\paragraph{} With regards to the rule determining which individual wins in a pairwise fight, two primary formulations have been applied: one in which the probability that the stronger individual wins depends on the difference in the strengths of the two individuals \cite{Bonabeau1995, Bonabeau1999, Odagaki2006, Odagaki2007, Albers2001, Boghosian2017}, and one in which this probability depends on a ratio of the strengths of the two individuals \cite{Dugatkin1997, Hemelrijk1999, Hogeweg1983}. We focus on the latter of these two formulations. This choice is related to our choice of the multiplicative rule for the amount of strength won or lost in the fight. In a multiplicative rule, large absolute differences in strength typically exist among individuals of similar rank, at the top-end of the strength distribution. Therefore upsets, in which the lower-strength individual defeats the higher-strength individual, become very unlikely or impossible at the top-end of the distribution of strength when the probability of winning depends on the difference in strengths of the two individuals. When the probability of winning depends on a ratio of the statuses of the two individuals, upsets tend to be more likely, especially between two high-strength individuals separated by a large absolute amount of strength.

\paragraph{} Conversely, in a model with an additive rule for the amount of strength won or lost, it may well be appropriate for the probability of winning to depend on the difference in strengths of the two individuals, since the status of an individual is equal to the difference in the number of times the individual has won and lost fights. However, especially in more complex animals, it is unrealistic to assume that the probability of winning is based on a tally of the number of fights won and lost, as this information is unavailable to the individuals involved in the fight. Rather, a more realistic assumption is that a psychological process occurs in which the two individuals make a rough comparison of one another's relative strengths, where this comparison influences each individual's probability of winning via characteristics such as confidence, willingness to take risks, and aggressiveness \cite{Chiao2010, Dugatkin1997}. This assumption is supported by psychological research showing that perceived change of a physical stimulus depends on the relative rather than the absolute change in the stimulus \cite{Hsu2006, Laming2009}.

\begin{figure}[H]
 \centering
 \includegraphics[scale=0.6]{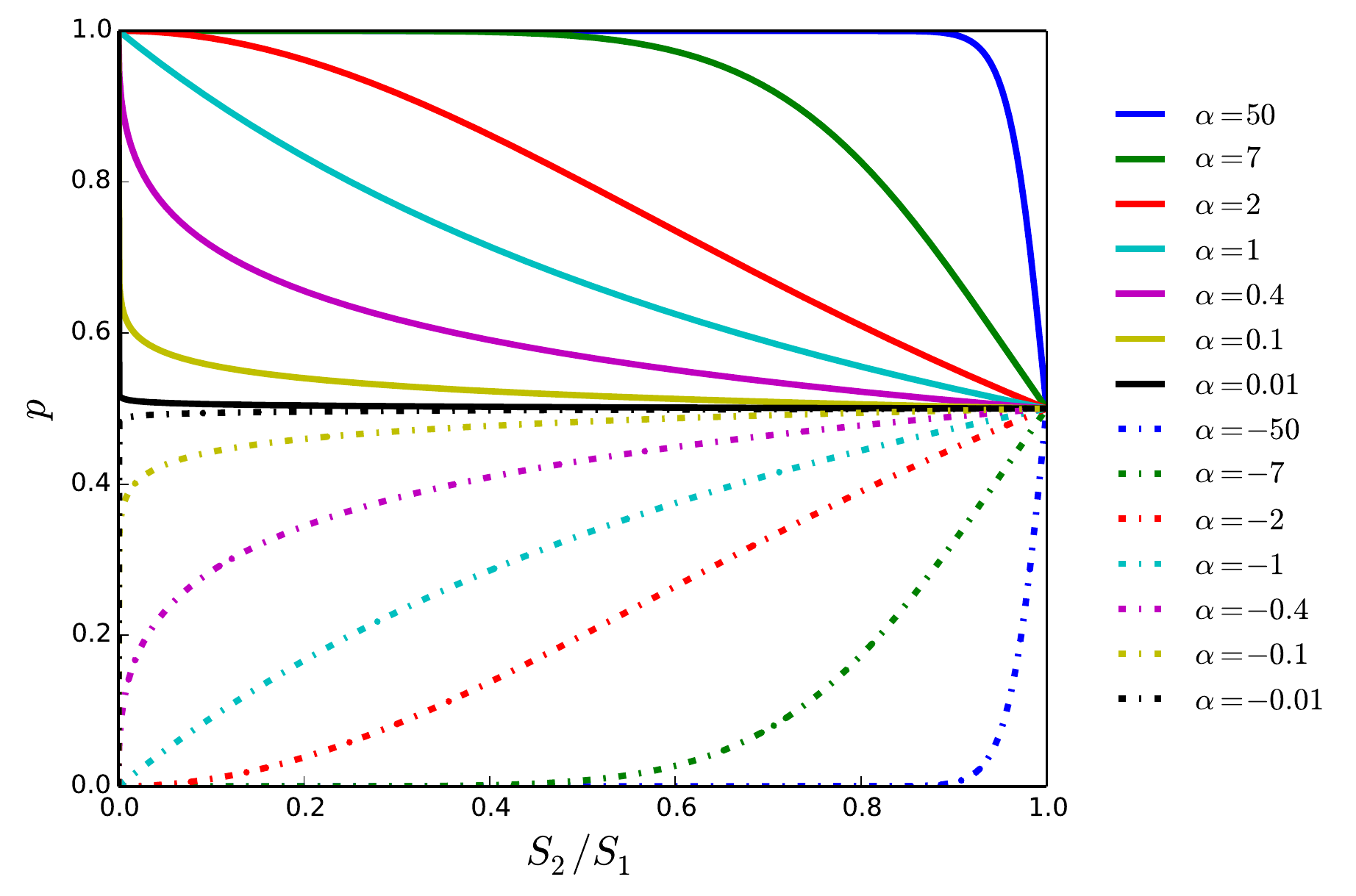} \caption{{\bf Probability that higher-status individual wins in a pairwise fight}. The probability $p(S_2/S_1)$ (Eq \ref{Eq:p}) is shown for different values of $\alpha$. Solid lines correspond to $\alpha>0$ and dash-dotted lines to $\alpha<0$.}
 \label{fig:p_v_x}
\end{figure}

\paragraph{} Our specific model is constructed as follows. We consider a system of $N$ individuals, each possessing a strength property, $S$, that determines the individual's societal position. We intend $S$ to represent the societal status of the individual, and accordingly we refer to ``status'' rather than the generic term ``strength'' in the remainder of this article. At each step in the simulation, a pair of individuals is randomly selected, and engages in a ``fight''. The probability, $p$, that the higher-status individual wins the fight is expressed as a function of its status, $S_1$, and that of its (lower or equal status) opponent, $S_2$:

\begin{equation}
\label{Eq:p}
 p = \frac{1}{1+(S_2/S_1)^\alpha}.
\end{equation}
When $\alpha=1$, the probability that either individual wins is equal to the ratio of its own status to the sum of its and its opponents statuses. However, as $\alpha$ is tuned to values other than $1$, the advantage held by the higher-status individual changes (Fig \ref{fig:p_v_x}). As $\alpha\to\infty$, the higher-status individual is virtually guaranteed to win, regardless of how strong its opponent is. On the other hand, when $\alpha$ is small but positive, the higher-status individual only has a large advantage in fights against opponents with much lower status. When $\alpha$ is negative, $0 \leq p < 0.5$, indicating that the lower-status individual in any given fight is more likely to win. The parameter $\alpha$ thus generalizes previous modeling approaches, by allowing the probability for winning a pairwise fight to be continuously adjusted between end-points where the lower-status individual always wins ($\alpha=-\infty$) and where the higher-status always wins ($\alpha=\infty$). 

\paragraph{} To interpret the societal meaning of the parameter $\alpha$, we note that the probability $p$ depends on the relative statuses of the two individuals. This means that as long as the ratio $S_2/S_1$ is constant, and given a constant value of $\alpha$, the probability, $p$, that the higher-status individual will win is constant, independent of the absolute values of $S_1$ and $S_2$. In a general sense, the probability that a high-status individual will win in a fight against a medium-status individual is the same as the probability that a medium-status individual will win in a fight against a low-status individual. If having a higher societal status can be considered as having a higher level of ``authority" in a hierarchical society, then the parameter  $\alpha$ represents the degree to which there is deference to authority in the society or, in other words, the society's overall level of ``authoritarianism". 

\paragraph{} Next, we explain the rule determining the amount of status transferred from loser to winner following each fight interaction. Let $S_W$ be the before-fight status of the winner of the fight, and $S_L$ the before-fight status of the loser. Following the fight, a portion $\Delta$ of the loser's before-fight status is transferred to the winner, such that

\begin{equation*}
\label{Eq:fight_outcome1}
\begin{aligned}
  S_W' &= S_W+\Delta \\
  S_L' &= S_L-\Delta,
\end{aligned}
\end{equation*}

where the primed quantities represent after-fight statuses. In our model, the amount of status transferred, $\Delta$, is equal to a proportion of the before-fight status of the individual who loses the fight. That is, $\Delta = \delta S_L$, where $\delta$ is a fraction between 0 and 1. This gives us: 

\begin{equation}
\label{Eq:fight_outcome2}
\begin{aligned}
  S_W' &= S_W+\delta S_L\\
  S_L' &= S_L-\delta S_L,
\end{aligned}
\end{equation}

\paragraph{} This rule for the amount of status transferred has realistic implications from the perspective of formation and maintenance of social hierarchy, because it means that upsets (in which the lower-status individual defeats the higher-status individual) produce large rewards for the winner and large penalties for the loser. 

\paragraph{} The two rules contained in Eqs \ref{Eq:p} and \ref{Eq:fight_outcome2} constitute our ``original" (two-parameter) model that is the main focus of this work. We note that special cases of Eq \ref{Eq:p} were investigated in previous work. Specifically, the case when $p=0.5$ in all fights, regardless of the statuses of the two individuals ($\alpha=0$ in our model) and the case when $p=1$ in all fights ($\alpha=\infty$ in our model) were investigated in a model that uses the same rule as our Eq \ref{Eq:fight_outcome2} \cite{Ispolatov1998}. The special case when $\alpha=1$ has been used in other winner-loser models, but never in conjunction with a multiplicative rule like our Eq \ref{Eq:fight_outcome2}. The novel parameter $\alpha$ thus allows us to generalize previous models and opens a previously unexplored region of parameter space. We also present a simple extension to our two-parameter model, in the following section.

\subsection{Model extension: restricting fights between individuals with large differences in status}
\label{section:extended_model}

\paragraph{} One of the main assumptions of winner-loser models based solely on the two categories of rules described in section \ref{section:the_model} is that any pair of individuals are equally likely to interact, regardless of their strengths. Some biologically-oriented winner-loser models have included mechanisms that adjust the interaction probability of individuals based on their spatial positions or on their strengths. For example, in Ref. \cite{Dugatkin1997}, each individual decides whether to engage in a fight by comparing the ratio of its strength to its opponent's strength with a threshold;  in Ref. \cite{Hemelrijk1999} individuals move in a spatial territory and interact if they are within visual range of one another; and in Ref. \cite{Bonabeau1999}, individuals interact with a probability equal to the product of a function of their strengths, such that stronger individuals interact more frequently than weaker individuals. In a similar vein, we can extend our model by implementing a third model rule under which pairs of individuals with large differences in status fight less often than similar-status individuals. Unlike other rules that control the probability that two individuals interact, our rule allows all individuals with similar statuses to interact frequently, while also reducing the frequency of interactions (and thus the exchange of status) between individuals with large differences in status. 

\paragraph{} These are needed realistic features, because evidence from studies of hierarchies in animal groups suggests that a large proportion of the status-determining interactions experienced by high status individuals pit these individuals against ``challengers'' who themselves have higher than average status \cite{Gesquiere2011, Sapolsky2011, Sapolsky1992}. Meanwhile, low status individuals tend not to challenge high status individuals, such that low status individuals are more likely to interact amongst themselves \cite{Sapolsky1992}. Similarly, humans are more likely to interact with members of their own social classes, especially at the extremes of the social class spectrum \cite{Cote2017}, and residential segregation, which impedes interactions between members of different social classes, is considered to be a primary factor in the creation and exacerbation of social stratification \cite{Massey1993}.

\paragraph{} Specifically, our extended model introduces two additional parameters. First, following from observations that high status individuals are more likely to engage in fights with similarly high status challengers, the new rule imposes that two selected individuals only engage in a fight if their absolute statuses are separated by not more than a threshold amount $\eta\bar{S}$. Here, $\eta \geq 0$ is a new parameter that sets the size of the threshold relative to the (conserved) average status of the system, $\bar{S}$, which is a natural reference point for the threshold position. Secondly, notwithstanding the above-noted observations regarding the higher frequency of interactions between similar-status individuals, animal behaviour studies also show that high status individuals do interact with low status individuals at times. This occurs, for example, through seemingly random acts of aggression which may play an important role  in maintaining hierarchical rank-ordering \cite{Silk2002}. For this reason, a realistic model should not exclude the possibility that fights between high and low status individuals will occasionally occur. Therefore, regardless of the result of the threshold criterion, the fight between the two selected individuals takes place if $r < \epsilon$, where $0 \leq \epsilon \leq 1$ is a new parameter and $r$ is a random number such that $0 \leq r < 1$. 

\paragraph{} In summation, the new rule to limit interactions based on the statuses of the two competitors can be stated as follows: two individuals are selected at random, and they fight if $S_1-S_2 \leq \eta\bar{S}$ OR $r \leq \epsilon$. We note that, in the implementation of the simulation, this rule does not change the probability with which any two particular individuals are selected from the population, but only adds a threshold criterion to decide whether or not the fight occurs between the selected pair.

\section{Time-evolution and structure of status distributions}
\label{section:struct_and_time_evol}

\paragraph{} In the society envisioned in the model, pairs of individuals interact such that they gain or lose societal status in accordance with the rules described in section \ref{section:the_model}. As interactions take place, and status is exchanged between the members of the society, a distribution of societal status takes shape. Our primary goal in this study is to investigate the structure of these status distributions and how they evolve in time. In this section, we therefore investigate the shapes of the status distributions as functions of the model parameters $\delta$ and $\alpha$ (section \ref{section:overview_status_distns}), and then quantify their time evolution in terms of two characteristic times (sections \ref{section:longlived_behaviour}-\ref{section:delta_alpha_diagram}). Before presenting these results, we first (section \ref{section:time}) establish how time is defined in the model. This introduces the first characteristic time of the system's evolution, which gives us a basis on which to present the results in the following sections. Please note that in the directly following sections we focus on the original (two-parameter) version of the model first as many features are qualitatively the same as for the extended version of the model. The features specific to the extended model are then discussed in section \ref{section:extended_model_distributions}.

\subsection{Definition of time and the characteristic time $\tau_1$}
\label{section:time}

\paragraph{} In order to discuss the shapes and time-evolution of the status distributions formed by simulations of the model, we must establish how time is defined. For simplicity and without loss of generality, we model the pairwise interactions as instantaneous such that they can be described as sequential events. We consider that one unit of time has passed once all members of the society have, on average, engaged in one pairwise interaction or fight. Under this definition of time, the rate at which an individual participates in a fight is an intrinsic frequency of the system, independent of system size, $N$, where $N$ is the number of individuals in the system. Time, $t$, is therefore defined as $t = 2t'/N$, where $t'$ is the number of fights that have occurred since the initiation of the simulation, and the factor of 2 comes from having each interaction involve two individuals. One unit of time is equal to $N/2$ fights.

\paragraph{} Previous work by Ispolatov et al. \cite{Ispolatov1998} shows the existence of a characteristic time in a model that is mathematically equivalent to our original (two-parameter) model in the case where $\alpha=0$. An analytic solution for the time evolution of the variance of the status distribution (wealth distribution, in Ref. \cite{Ispolatov1998}) was found to be as follows: 

\begin{equation}
 \label{Eq:M2_alpha0}
M_2(t) = \frac{\delta\bar{S}}{1-\delta} \left(1-e^{-\delta(1-\delta)t}\right),
\end{equation}

where the variance (second central moment) $M_2(t)$  can be calculated directly from the status distribution: 

\begin{equation}
 \label{Eq:M2_def}
M_2(t)=\frac{\sum_{i=1}^{N}{\left(S_i(t)-\bar{S}\right)^2}}{N},
\end{equation}

where $S_i(t)$ is the status of individual $i$ at time $t$, and $\bar{S}$ is the (conserved) average status. 

\paragraph{} Similar to Eq \ref{Eq:M2_alpha0}, higher moments of the status distribution converge to constant values, and the status distribution attains a steady state. Eq \ref{Eq:M2_alpha0} therefore shows that the variance approaches a steady-state value of $M_2=\delta\bar{S}/(1-\delta)$ at large times and that the approach to the steady-state is characterized by a time constant equal to $(\delta(1-\delta))^{-1}$. Eq \ref{Eq:M2_alpha0} can be re-written in a form that is useful for our purposes:

\begin{equation} 
 \label{Eq:exp_c1}
 M_2 = c_1(1-e^{-t/\tau_1}),
\end{equation}

where $c_1$ and $\tau_1$ are generally functions of $\delta$ and $\alpha$. In the following, we use the symbols $\hat{c_1}$ and $\hat{\tau_1}$ to represent these functions when $\alpha=0$, such that  $\hat{\tau_1}=(\delta(1-\delta))^{-1}$ and $\hat{c_1}=\delta\bar{S}/(1-\delta)$, as per Eq \ref{Eq:M2_alpha0}. \nameref{S1_Appendix} (part A) contains a demonstration that our definition of time corresponds to how time is defined in the analytical result in Eq \ref{Eq:M2_alpha0}.

\subsection{Overview of status distributions produced by the model}
\label{section:overview_status_distns}

\paragraph{} In this section, we present the shapes of the societal structures (distributions of status) that emerge in simulations of the model. We begin by setting $\alpha=0$ ($p=0.5$ for all fights, regardless of the statuses of the competitors as per Eq \ref{Eq:p}) because, in this limiting case, there always results a stable steady-state status distribution as we show in the following.

\begin{figure}[H]
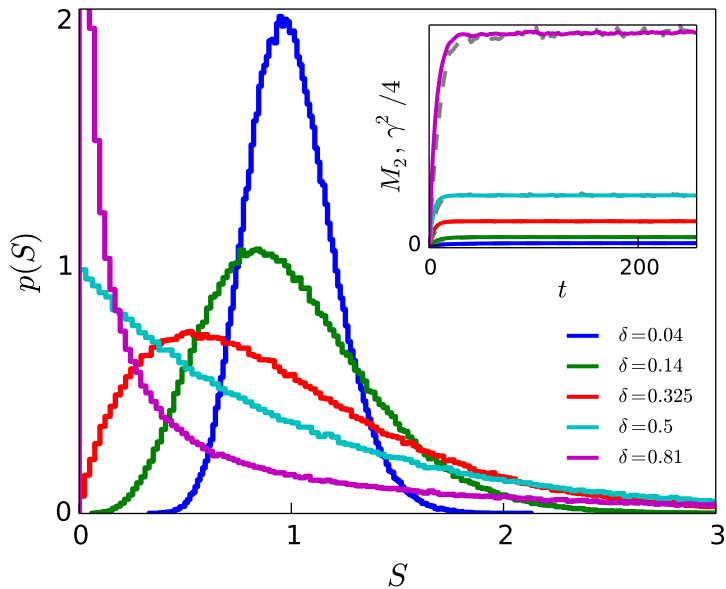

 \centering
 \includegraphics[scale=1.0]{{{PSvS_N100000_nr5_alpha0_single_panel}}}
 \caption{{\bf Shape of status distributions as function of $\delta$, with $\alpha=0$.} Distributions range from more egalitarian (e.g. $\delta=0.04$: all individuals have close to the average status) to highly unequal (e.g. $\delta=0.81$: most individuals have very low status and small portion of the population has very high status). Inset of (a): plateau in $M_2$ (coloured) and squared skewness (grey) indicate that status distributions are in steady-state. Larger values of $\delta$ correspond to larger steady-state value of $M_2$, such that $M_2$ provides a measure of the level of inequality of the society. Distributions obtained after simulating up to time $t=64\hat{\tau_1}$. $N=10^5$, $n_r=5$.}
 \label{fig:PSvS_alpha0}
\end{figure}

\paragraph{} In Fig \ref{fig:PSvS_alpha0}, we show graphs of steady-state distributions for several values of $\delta$ when $\alpha=0$. As can be seen, the shape of the steady-state distribution varies from rather egalitarian for small $\delta$ (e.g. $\delta=0.04$: all individuals have close to the average status) to highly unequal (e.g. $\delta=0.81$: most individuals have very low status and small portion of the population has very high status). As was noted in section \ref{section:time}, when $\alpha=0$, our model is mathematically equivalent to the model of Ispolatov et al. \cite{Ispolatov1998}, who showed that the tail of the distribution decays exponentially for all values of $\delta$.

\paragraph{} The inset of Fig \ref{fig:PSvS_alpha0} shows the time evolution of the variance of the status distribution, $M_2$, which provides a measure of the level of inequality of the society. Larger values of $\delta$ give rise to larger steady-state value of $M_2$. As expected from Eq \ref{Eq:M2_alpha0}, $M_2$ approaches a steady-state plateau with a value of  $\hat{c_1}$.  The skewness $\gamma = M_3/(M_2)^{3/2}$, where $M_3$ is the third central moment of the distribution, also arrives at a steady-state plateau at large time (dashed grey lines in inset of Fig \ref{fig:PSvS_alpha0}). This plateau in $M_2$ and $\gamma$ indicates that the shape of the status distribution is unchanging in time. The plateau in $\gamma^2$ is equal to four times that of the plateau in $M_2$, as can be shown by solving for the third moment following the approach presented in Ref. \cite{Ispolatov1998}.

\begin{figure}[H]
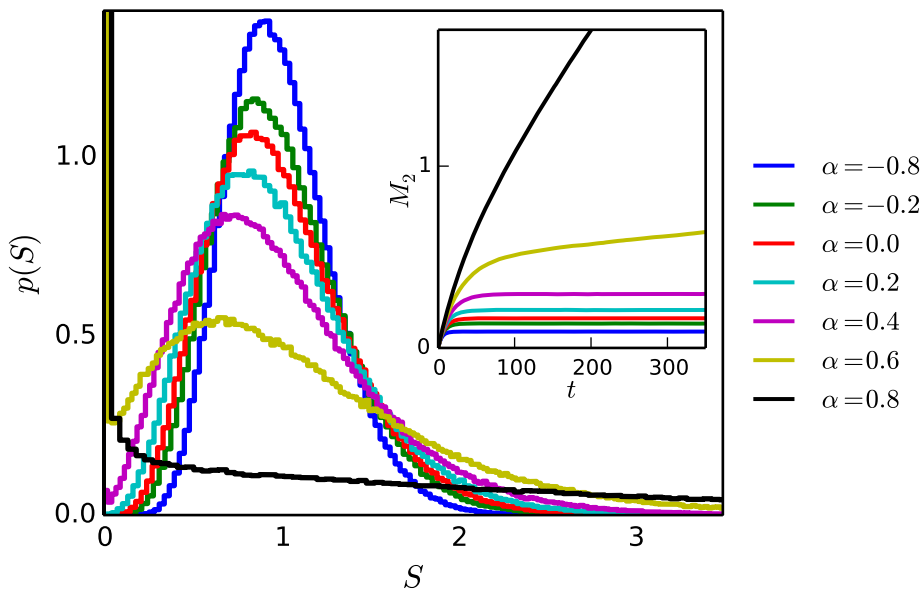

 \centering
 \includegraphics[scale=1.0]{{{PSvS_N100000_nr5_alpha_neq_0_single_panel}}}
 \caption{{\bf Shape of status distributions as functions of $\alpha$, with $\delta=0.14$.} Increasing $\alpha$ leads to an increase in the level of inequality of the society, while decreasing $\alpha$ leads to a decrease in the level of inequality. Inset: when $\alpha>0$, $M_2$ does not attain a plateau but continues to increase with $t$, at a rate that depends on $\alpha$. The status distribution appears to be in steady-state for small values of $\alpha>0$ (e.g. $\alpha=0.2$ and $\alpha=0.4$ curves in the inset) when observed on short enough time scales, while they are in fact not. For larger values of $\alpha>0$ (e.g. $\alpha=0.6$ and $\alpha=0.8$ curves in the inset), the level of inequality noticeably increases on the observation timescale, indicating a runaway of the status distribution toward an end-state of maximum inequality. Distributions obtained after simulating up to time $t=64\hat{\tau_1}$. $N=10^5$, $n_r=5$.}
 \label{fig:PSvS_alpha_neq_0}
\end{figure}

\paragraph{} Fig \ref{fig:PSvS_alpha_neq_0} shows distributions of societal status obtained for a fixed value of $\delta$ and for various values of the authoritarianism $\alpha$. The curve for $\alpha=0$, $\delta=0.14$ from Fig \ref{fig:PSvS_alpha0} is reproduced in Fig \ref{fig:PSvS_alpha_neq_0}, along with the inset, showing that $M_2(t)$ undergoes an initial transient period before arriving at a plateau value for times $t \gg \hat{\tau_1}$. 

\paragraph{} However, for large values of $\alpha>0$ ($\alpha=0.6$ and $\alpha=0.8$), the inset of Fig \ref{fig:PSvS_alpha_neq_0} shows a rapid increase of $M_2(t)$ that continues beyond the initial transient period. The status distributions are rapidly evolving (``running away'') toward a totalitarian end-state in which a single individual possesses virtually all of the societal status of the system and all other individuals have status approaching zero. The shape of the status distribution changes rapidly during this evolution, becoming more and more skewed with time. Additional figures showing how the status distributions evolve over long times are contained in Part E of \nameref{S1_Appendix}, and part F of \nameref{S1_Appendix} contains a proof that only one individual with non-negligible status remains in the totalitarian end-state.

\paragraph{} For smaller values of $\alpha>0$ ($\alpha=0.2$ and $\alpha=0.4$), $M_2(t)$ (and therefore the shape of the status distribution) appears to be virtually unchanged in the time following the initial transient period shown in Fig \ref{fig:PSvS_alpha_neq_0}. Yet, as we show below (sections \ref{section:longlived_behaviour} and \ref{section:phenom_tau2}), $M_2(t)$ does increase with $t$, albeit much more slowly, such that the status distributions can be considered to be in a long-lived state, where the shape of the distribution can change so slowly that it is essentially unchanged over sufficiently short observation times.

\paragraph{} Fig \ref{fig:PSvS_alpha_neq_0} also shows the status distributions that arise when $\alpha < 0$. For this region of parameter-space, the lower-status individual has a higher probability of winning the fight than the higher-status individual. Our numerical simulations show that the distributions are in steady-state (as for $\alpha=0$) and become more egalitarian (smaller $M_2$) as $\alpha$ is decreased while $\delta$ is held constant.

\paragraph{} The plots in Figs \ref{fig:PSvS_alpha0} and \ref{fig:PSvS_alpha_neq_0} were obtained using an ``egalitarian'' initial condition, in which all individuals have an initial status of $S=1$. \nameref{S1_Appendix} (part B) contains figures showing that the same distributions arise when the system is prepared in a ``uniform'' initial condition in which the initial statuses are randomly selected from a uniform distribution with $\bar{S}=1$.

\subsection{Long-lived behaviour}
\label{section:longlived_behaviour}

\begin{figure}[H]
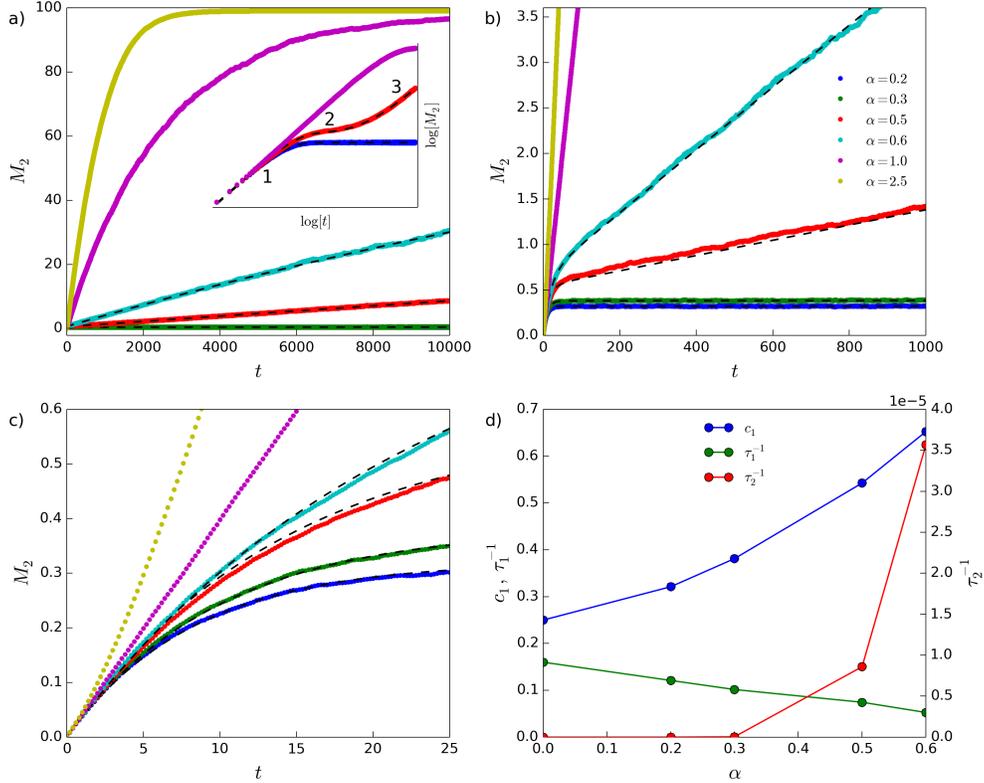

 \centering
 \includegraphics[width=\linewidth]
 {{{sum_exp_fit_with_upper_plateau_delta0.2_N100_partd_params_parta_logloginset}}}
 \caption{{\bf Evolution of $M_2(t)$ for $\alpha>0$.} Values of $\alpha$ are indicated in the legend in (b), and $\delta=0.2$ for all curves. (a) shows rapid ascent of $M_2(t)$ to upper plateau value for $\alpha=1.0$ and $\alpha=2.5$ (main plot), and the inset of (a) shows three of the curves on log-log scale, with stages 1-3 of the evolution of $M_2(t)$ indicated for the $\alpha=0.5$ curve. (b) and (c) show the main plot of (a) at different magnifications, to allow inspection of the fit of Eq \ref{Eq:sum_exp} (dashed black lines) to the curves with $\alpha \leq 0.6$. (d) fit parameters, with y axis scale for $\tau_2^{-1}$ shown on righthand side (parameters for $\alpha=0$ are from Eq \ref{Eq:sum_exp}, assuming $\tau_2=\infty$). $N=100$, $n_r=500$.}
  \label{fig:M2_v_t_with_fits}
 \end{figure}

\paragraph{} In this section we quantify the time evolution and the long-lived behaviour of the status distributions produced by the model by examining the evolution of the variance of the status distribution, $M_2(t)$, when $\alpha>0$ (Fig \ref{fig:M2_v_t_with_fits}a-c).

\paragraph{} As noted in section \ref{section:overview_status_distns}, for large values of $\alpha$, the status distribution runs away to an end-state in which a single individual possesses virtually all of the status in the society, and all other individuals have virtually zero status. In this totalitarian end-state, the variance $M_2$ approaches an upper plateau 
\begin{equation} 
\label{Eq:c2}
  c_2 = (N-1)\bar{S}^2=N-1,
\end{equation}
where the average status is defined in the model to be $\bar{S}=1$, without loss of generality. The upper plateau $c_2$ is the maximum possible value of $M_2$ (indicating the maximum level of inequality of the society). For finite-sized systems, $M_2$ rises to this upper plateau at large times. This large-time ascent to $c_2$ happens quickly for large values of $\alpha$, and much more slowly for smaller values of $\alpha$ (main plot of Fig \ref{fig:M2_v_t_with_fits}a).

\paragraph{} There therefore appear to be two relevant time-scales in the dynamics of the status distributions: one controlling the evolution away from the initial condition, and a second controlling the long-time approach to the totalitarian end-state. To capture the dynamics of $M_2(t)$ for $\alpha>0$, we attempt to fit, to the simulation data, a sum of exponential functions: 
\begin{equation} 
 \label{Eq:sum_exp}
 M_2 = c_1(1-e^{-t/\tau_1})+(c_2-c_1)(1-e^{-t/\tau_2}),
\end{equation}
where $\tau_2$ is a characteristic time controlling the rate of approach of $M_2$ to the upper plateau. The first term in Eq \ref{Eq:sum_exp} relates to the short-time dynamics of the status distribution, while the second term relates to the long-time dynamics. Long-lived states are produced for values of the model parameters $\alpha$ and $\delta$ for which $\tau_2$ is much larger than the time, $\tau_{obs}$, over which the system is observed (simulated), and $\tau_1$. When $\tau_1 \ll t \ll \tau_2$, Eq. \ref{Eq:sum_exp} becomes $M_2(t) \approx c_1(1-e^{-t/\tau_1})$ where, for $\alpha>0$, $c_1$ represents an operational plateau value of $M_2(t)$ corresponding to the long-lived state.

\paragraph{} Fits of Eq \ref{Eq:sum_exp} to simulated data are shown by the black dashed lines in Fig \ref{fig:M2_v_t_with_fits}. For these fits, $c_2$ is fixed at its upper plateau value $c_2=N-1$. Although the fits are imperfect (Fig \ref{fig:M2_v_t_with_fits}c), they do appear to capture the short-time ``elbow" controlled by $\tau_1$ and $c_1$, and the long-time ascent toward the upper plateau controlled by $\tau_2$ (Fig \ref{fig:M2_v_t_with_fits}b). Fig \ref{fig:M2_v_t_with_fits}d shows the fit parameters as functions of $\alpha$: as $\alpha$ is increased, both $c_1$ and $\tau_1$ increase, resulting in a slower evolution of $M_2$ (and therefore, of the shape of the status distribution) away from the initial condition. On the other hand, as $\alpha$ increases, $\tau_2$ decreases, leading to a more rapid (long-time) approach to the upper plateau. As $\alpha$ is increased to larger values, it becomes difficult to resolve the early-time ``elbow", and thus difficult to obtain a meaningful fit of Eq \ref{Eq:sum_exp}. Furthermore, as shown in Fig \ref{fig:M2_v_t_with_fits}c, the shape of $M_2(t)$ appears to approach a straight line (at early times) as $\alpha \to 1$, and then, to bend upwards away from such an initial straight line when $\alpha > 1$. 

\paragraph{} For small $\alpha>0$, the status distributions pass through three phenomenological stages, beginning with an egalitarian initial condition and ending in the totalitarian end-state (see \nameref{S1_Appendix}, part E for details). The first stage pertains to the evolution of the status distribution away from its initial condition over time scales of the order $\tau_1$ and into a distribution with a form similar to that of the $\alpha=0$ steady-state distribution. This ``stage 2'' distribution changes only very slowly, eventually transitioning into a stage (``stage 3") where high status individuals are nearly guaranteed to win all fights. The duration of stage 2 decreases and essentially disappears as $\alpha$ is increased, explaining the inability of Eq \ref{Eq:sum_exp} to represent $M_2(t)$ for larger values of $\alpha$. In the asymptotic state of the evolution $t \gg \tau_2$, all individuals have $S \approx 0$, except for a single individual with status equal to the total status of the system.  

\paragraph{} Our findings are largely independent of the system size $N$. As shown explicitly in the SI (\nameref{S1_Appendix}, part C), the parameters controlling the early-time behaviour of $M_2$ ($c_1$ and $\tau_1$) remain constant as the system size, $N$, is increased, whereas the parameters controlling the long-time behaviour of $M_2$ ($c_2$ and $\tau_2$) both scale linearly with $N$. A proof that the time required to reach the end-state, $\tau_{end}$, scales linearly with $N$ in the extreme scenario where $\delta=1$ and $\alpha=\infty$ is also included in \nameref{S1_Appendix} (part D), as a demonstration of the configurational reasons why the long-time (approach to the end-state) evolution of the model dynamics increases in proportion to the system size $N$.

\subsection{Phenomenology of the characteristic time $\tau_2$ when $\alpha>0$}
\label{section:phenom_tau2} 

\paragraph{} The characteristic time $\tau_2$ controls the rate at which the system approaches the totalitarian end-state when $\alpha>0$. This characteristic time increases as $\alpha$ is decreased from large positive values, as shown in Fig \ref{fig:M2_v_t_with_fits}d. We can also see, from comparison of Eqs \ref{Eq:M2_alpha0} and \ref{Eq:sum_exp}, that $\tau_2=\infty$ when $\alpha=0$. We would like to know the functional relationship between $\tau_2$ and the model parameters in order to quantitatively characterize the transition between long-lived and runaway behaviours. To further explore this relationship, we consider an analogy with the barrier-like or  ``activated'' processes typical of many physical systems \cite{Hanggi1990}. We find that the resulting Arrhenius equation provides a good description of the relationship between $\tau_2$ and the model parameters $\alpha$ and $\delta$. This allows us to determine, as a function of the model parameters, the observation times over which status distributions can be considered long-lived, which we summarize in the following section.

\paragraph{} The rate of an activated process is proportional to an exponential term containing an energy barrier scaled by temperature. The exponential term is multiplied by a pre-factor called the attempt frequency, which is typically independent of temperature. This type of relationship between rate and temperature can be found in many diverse physical phenomena, including the rate of chemical reactions \cite{Hanggi1990}, the relationship between diffusion coefficients and temperature \cite{Hanggi1990}, the rate of nucleation according to the classical nucleation theory \cite{Markov2003}, the viscosity of strong glass-formers \cite{Angell1991}, and the blocking transition in superparamagnetism \cite{Knobel2008}, as well as in biology regarding, for example, the rate of chirping in crickets and of flashing in fireflies, and in psychology, where human perception of time is related to body temperature through a relationship of this form \cite{Laidler1972}.

\paragraph{} If the characteristic time $\tau_2$ is regulated by $\alpha$ according to an activated process, then one would expect the relationship between $\tau_2$ and $\alpha$ to follow an Arrhenius equation of the form: 

\begin{equation}
\label{Eq:Arrhenius}
\frac{1}{\tau_2} = f_0 e^{-\alpha_b/\alpha},
\end{equation}

\begin{figure}[H]
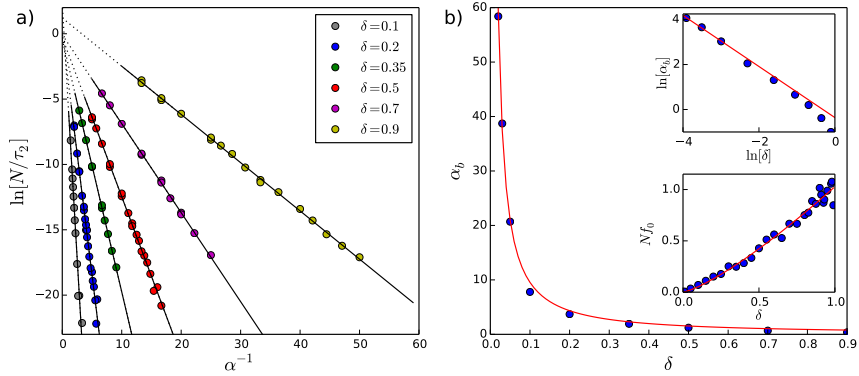

 \centering
 \includegraphics[width=\linewidth]
 {{{multipanel_Arrhenius}}}
 \caption{{\bf Arrhenius relationship between $\tau_2$ and $\alpha$ and $\delta$.} a) Plots of $\ln[N/\tau_2]$ vs. $\alpha^{-1}$ for various values of $\delta$ confirm the relationship proposed in the Arrhenius equation (Eq \ref{Eq:Arrhenius}). The slope of each linear fit is $-\alpha_b(\delta)$. A discussion regarding the evaluation of errors on the extracted values of $\tau_2$ is included in \nameref{S1_Appendix} (part G). b) Dependence of $\alpha_b$ and $Nf_0$ on the parameter $\delta$: the red line in the main plot (linear scale) and upper inset (logarithmic scale) corresponds to $\alpha_b = 0.53\delta^{-1.21}$; the red line in the lower inset corresponds to $Nf_0=1.03\delta^{1.28}$.}
 \label{fig:Arrhenius}
\end{figure}

\paragraph{} where $\alpha_b$ is a term that plays a role similar to an energy barrier in an activated process and $f_0$ is analogous to an attempt frequency. To test this idea, we plot the logarithm of $N/\tau_2$ vs. $\alpha^{-1}$ in Fig \ref{fig:Arrhenius}a. The factor of $N$ is included due to the fact that $\tau_2$ scales linearly with $N$, as discussed in section \ref{section:longlived_behaviour} and shown in \nameref{S1_Appendix} (part C).

\paragraph{} The linear behaviour seen in Fig \ref{fig:Arrhenius}a confirms the relationship between $\tau_2$ and $\alpha$ proposed in the Arrhenius equation (Eq \ref{Eq:Arrhenius}). In the figure, the $\delta$-dependent slopes of the linear fits correspond to $-\alpha_b$, and the intercepts to $\ln[Nf_0]$. The values of $\alpha_b$ extracted from the linear fits in Fig \ref{fig:Arrhenius}a are shown as a function of $\delta$ in Fig \ref{fig:Arrhenius}b. As can be seen, $\alpha_b$ diverges as $\delta$ is decreased. The red line in Fig \ref{fig:Arrhenius}b (main plot and upper inset) shows the function $\alpha_b = 0.53\delta^{-1.21}$.

\paragraph{} In Fig \ref{fig:Arrhenius}a, the y-intercepts of the linear fits appear to cluster around 0, suggesting that the prefactor $Nf_0$ in the expression for $N/\tau_2$ following from the Arrhenius equation is of the order of $1$ for the values of $\delta$ considered. The y-intercepts do not, however, give a robust determination of the prefactor $Nf_0$. This may be due to a change in functional form of Eq \ref{Eq:sum_exp} as $\alpha$ increases such that $\tau_1$ vanishes. Alternatively, the prefactor $Nf_0$ can be directly determined by setting $\alpha = \infty$ (equivalent to $p=1$ in Eq \ref{Eq:p}) in the simulations and extracting $\tau_2(\alpha=\infty)$ from $M_2(t)$. In so doing, we have assumed that $t \gg \tau_1$ and $c_2 \gg c_1$ such that Eq \ref{Eq:sum_exp} becomes $M_2(t) \approx c_2(1-e^{-t/\tau_2})$. $Nf_0$ is then equal to $N/\tau_2(\alpha=\infty)$. This approach provides more robust determinations of $Nf_0$, which are shown in the lower inset of Fig \ref{fig:Arrhenius}b, along with a fit (red line) of the function $Nf_0 = 1.03\delta^{1.28}$.

\paragraph{} Substituting the expressions for $Nf_0$ and $\alpha_b$ into the Arrhenius equation (Eq \ref{Eq:Arrhenius}) gives: 

\begin{equation}
\label{Eq:Arrhenius_subst}
\frac{N}{\tau_2} \approx 1.03\delta^{1.28}e^{-0.53/(\alpha\delta^{1.21})}.
\end{equation}

The Arrhenius equation and Eq \ref{Eq:Arrhenius_subst} show that $\tau_2$ diverges exponentially as $\alpha \to 0$. This is consistent with Eq \ref{Eq:M2_alpha0}, which indicates that $\tau_2 = \infty$ when $\alpha=0$. Moreover, the parameter $\delta$ appears in both the pre-factor and the argument of the exponential of Eq \ref{Eq:Arrhenius_subst}, such that as $\delta \to 0$, $\tau_2 \to \infty$, regardless of the value of $\alpha$. This is consistent with the fact that for $\delta = 0$ no status is exchanged during a pairwise interaction and the initial status distribution is trivially stable. These observations strongly suggest that stable status distributions only exist for $\alpha \leq 0$ as well as $\delta = 0$. When $\alpha > 0$, Eq \ref{Eq:Arrhenius_subst} allows us to identify the time scale of the transition between long-lived societal structures and runaway toward the totalitarian end-state as we discuss in more detail below.

\subsection{$\delta-\alpha$ phase diagram} 
\label{section:delta_alpha_diagram}

\begin{figure}[H]
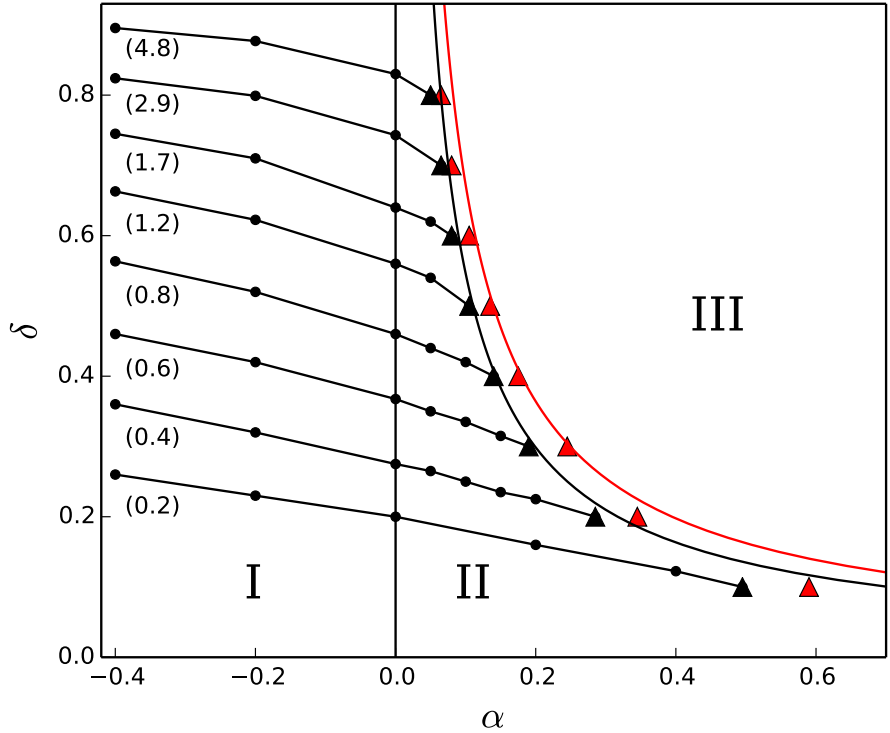

 \centering
 \includegraphics[width=\textwidth]
 {{{stability_diagram}}}
 \caption{{\bf $\delta-\alpha$ phase diagram.} The model exhibits three regions of behaviour in $\delta-\alpha$ parameter-space: I ($\alpha \leq 0$ or $\delta = 0$): true (infinite-duration) steady-state status distributions; II (small values of $\alpha > 0$): long-lived status distributions; III (large values of $\alpha>0$): runaway behaviour. Within regions I and II, equi-$M_2$ lines (lines of equal standard deviation of the status distribution) are shown with $M_2$ values indicated in parentheses below each equi-$M_2$ line. The location of the transition between regions II and III is observation time-dependent, and is marked by the black triangular points for $\tau_{obs}=10^4$ and by the red triangular points for $\tau_{obs}=10^3$, as determined directly from the simulation data. The location of the transition as determined by the Arrhenius relation between $\tau_2$ and $\alpha$ is shown by the black ($\tau_{obs}=10^4$) and red ($\tau_{obs}=10^3$) curves. System size $N=1000$.}
 \label{fig:stability_diagram}
\end{figure}

\paragraph{} As shown in sections \ref{section:overview_status_distns}-\ref{section:phenom_tau2}, for all positive values of $\delta$, the model gives rise to three regions of behaviour: true steady-state status distributions ($\alpha \leq 0$), long-lived status distributions that eventually arrive at the totalitarian end-state (small values of $\alpha > 0$), and runaway (rapid evolution) toward the totalitarian end-state (large values of $\alpha > 0$). These three regions and the transitions between are portrayed in a $\delta-\alpha$ phase diagram in Fig \ref{fig:stability_diagram}.

\paragraph{} Fig \ref{fig:stability_diagram} is a summary of the main results of our model. In it, we see the three regions of behaviour described in sections \ref{section:overview_status_distns}-\ref{section:phenom_tau2},. The region (marked with a roman numeral I) of infinite-duration steady-state status distributions is separated from a region (II) of long-lived status distributions by a transition that occurs due to an exponential divergence of the characteristic time $\tau_2$ as $\alpha \to 0^{+}$. Runaway behaviour (region III) occurs when a noticeable slope is observed in the evolution of $M_2(t0$ (see the inset of Fig \ref{fig:PSvS_alpha_neq_0} for $\alpha=0.6$ and $\alpha=0.8$). Whether such a slope is observed or not depends on $\tau_2(\delta, \alpha)$ and on the time, $\tau_{obs}$, over which the system is observed. The location, in $\delta-\alpha$ parameter-space, of the transition between regions II and III therefore depends on $\tau_{obs}$, and can be determined directly from the data or via Eq \ref{Eq:Arrhenius_subst}, as described below.

\paragraph{} We determined the location of the transition between regions II and III in two ways. First, we used a simple criterion to equate the onset of runaway with the appearance of a positive slope in the long-time portion of $M_2(t)$. In this way, the long-lived state corresponds to a plateau value of $M_2(t)$ over a particular $\tau_{obs}$ (where $\tau_{obs} \gg \tau_1$). The long-lived state is considered to be lost when, instead of a plateau, a positive slope is observed in $M_2(t)$ after a time $\tau_{obs}$ has transpired. The black and red triangular markers in Fig \ref{fig:stability_diagram} indicate the onset of runaway as determined by this criterion, for $\tau_{obs}=10,000$ and $\tau_{obs}=1,000$, respectively. Secondly, the location of the transition can be determined using the Arrhenius relation presented in section \ref{section:phenom_tau2} to determine the value of $\alpha$ for which $M_2(t)$ increases by a sufficient amount after a time $\tau_{obs}$ has transpired. The location of the transition as determined by the Arrhenius relation is shown by the curving curving black and red lines in Fig \ref{fig:stability_diagram}. Details about how these two approaches were conducted are contained in \nameref{S1_Appendix} (part I).

\paragraph{} While $\alpha$ largely determines the stability of the asymptotic status distributions, the exact value of $\delta>0$ influences the shape of the status distribution, as already seen for $\alpha = 0$ in section \ref{section:overview_status_distns}. This is also true for the long-lived distributions. Essentially identical distributions can be produced (within a particular simulation time) for different combinations of the parameters $\delta$ and $\alpha$ (see part H of \nameref{S1_Appendix} for details). Sets of such points are plotted as ``equi-$M_2$'' lines (lines of equal standard deviation of the status distribution) within regions I and II in Fig \ref{fig:stability_diagram}.

\subsection{Status distributions in the extended model}
\label{section:extended_model_distributions}

\paragraph{} In section \ref{section:extended_model}, we presented a simple extension to the original (two-parameter) model that restricts which individuals can fight each other based on the proximity in their statuses. This ``extended model'' introduces two new parameters: $\eta$, which sets a threshold  $\eta\bar{S}$ (where $\bar{S}$ is the average status of the system), such that individuals with statuses separated by an amount more than this threshold are restricted from fighting with each other; and $\epsilon$, the probability with which two individuals that have a separation of statuses greater than $\eta\bar{S}$ do nevertheless fight each other.

\begin{figure}[H]
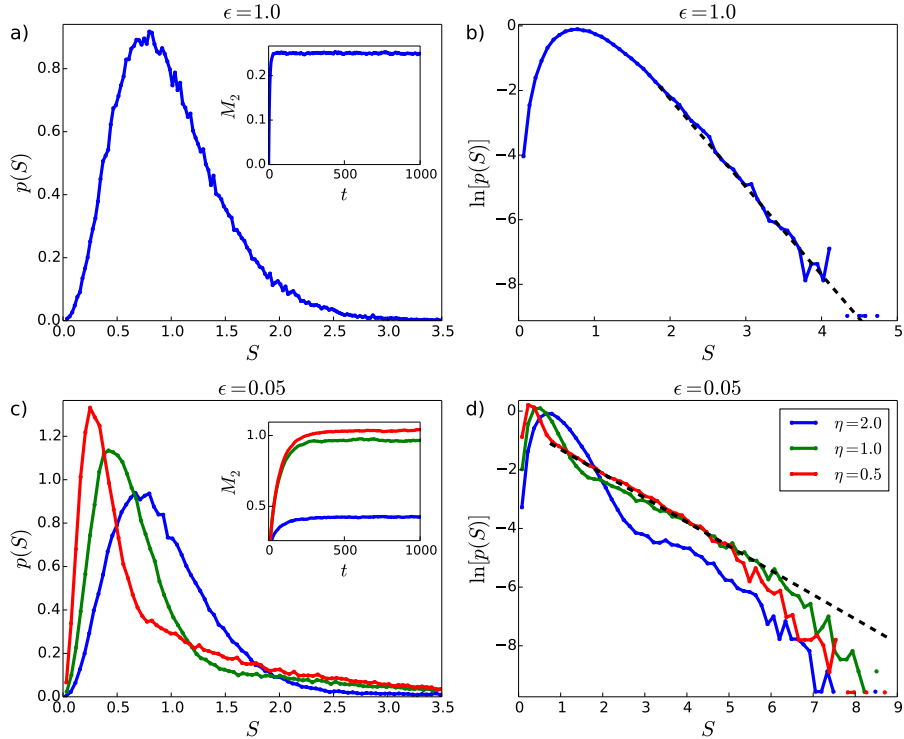
 
 \centering
 \includegraphics[scale=0.4]
 {{{histograms_eps_eta_composite_difference_model_abcd}}}
 \caption{{\bf Status distributions in the extended model.} $\delta=0.2$ and $\alpha=0$ in all plots. Plots (b) and (d) show the distributions on a logarithmic scale, in order to allow for inspection of the large-$S$ tail. When $\epsilon$ is decreased from $1$, a ``break'' in the distribution emerges, (particularly evident on the logarithmic scale) corresponding to a society with distinguishable low status and high status groups or classes. The black dashed lines in (b) and (d) are maximum likelihood fits of exponential distributions with lower-bound at $S=2.25$ in (b) and [lower-bound, upper-bound] at $S=[0.75,5.0]$ in (d), where plotted fit line in (d) is extrapolated beyond $S=5.0$. System size $N=10^5$.}
 \label{fig:histograms_eps_eta}
\end{figure}

\paragraph{} Fig \ref{fig:histograms_eps_eta} shows distributions of status generated by simulations of the extended model. A log-linear scale is used in the righthand column of plots to allow inspection of the high-status tails of the status distributions. In all subplots, $\alpha=0$ and $\delta=0.2$. When $\epsilon=1$ (Fig \ref{fig:histograms_eps_eta}a-b) the original (two-parameter) model is recovered, such that the parameter $\eta$ has no effect on the shape of the status distribution. Also, when $\epsilon=1$, the high-status tail of the distribution decays exponentially, in accordance with the analytic result found by Ispolatov et al. \cite{Ispolatov1998} (Fig \ref{fig:histograms_eps_eta}b). When $\epsilon$ is decreased, a ``break" in the high-status tail of the distribution emerges, as can be seen in the log-linear plot in Fig \ref{fig:histograms_eps_eta}d. Following this break, the distribution enters a second exponentially-decaying regime (black dashed line in Fig \ref{fig:histograms_eps_eta}d) that ends with a cutoff. Increases to $\eta$ cause the location of the break as well as the location of the peak of the distribution to shift to higher values of $S$. The plateaus in $M_2(t)$ (insets in Fig \ref{fig:histograms_eps_eta}) for $\epsilon>0$ indicate that these distributions are in steady-state.

\paragraph{} When $\epsilon = 0$ (not shown), $M_2(t)$ does not obtain a plateau and continues to increase over the duration of the simulation time. For this value of the parameter $\epsilon$, the system approaches an end-state in which the majority of individuals have status approaching zero, and a small minority of individuals have large statuses. In this end-state, the few high-status individuals are prevented from interacting with each other because their statuses are separated by amounts greater than $\eta\bar{S}$. The specific configuration of the $\epsilon=0$ end-state depends on the particular sequence of interactions. A positive value of $\epsilon$ is therefore needed in order for the simulation to obtain a unique steady-state.

\paragraph{} In the plots in Fig \ref{fig:histograms_eps_eta}, $\alpha=0$. The distributions produced with $\alpha=0$, $\eta>0$, and $\epsilon>0$ show a plateau in $M_2(t)$. When $\alpha>0$, $M_2(t)$ behaves qualitatively in the same way as the original (two-parameter) model. That is, a long-lived state is observed for small values of $\alpha>0$, and a runaway for large values of $\alpha>0$, where the location of the transition between the long-lived state and runaway depends on the observation (simulation) time. Several figures showing how the extended model distributions evolve for representative values of $\alpha>0$ are included in \nameref{S1_Appendix} (part J).

\section{Comparison of model results with real-world data} \label{section:realworld}

\paragraph{} In this section, we compare the results of our model to data from real-world social hierarchies in two ways. First, we consider data on agonistic interactions (fights) from animal observation studies, in section \ref{section:agonistic}. We are able to make some general comments about the parameter values and the stage in the time evolution of the system for which the win and loss patterns in the simulations resemble and are consistent with the animal behaviour data. We note that the available data in this case is for small system sizes. Thus, the history of each particular pair of individuals' past interactions might be important since the individuals do recognize and remember one another. This feature is not captured by our model, which is a better description for large groups of individuals, in which there is a large probability that an individual $i$'s next interaction will be with an opponent with whom $i$ did not interact recently.

\paragraph{} Unfortunately, there are currently no observational interaction data for sufficiently large groups of individuals. In order to compare the distributions of societal status from the model with real-world social hierarchies, we therefore seek a measurable quantity that serves as a proxy for societal status in large social groups. Such a proxy must allow the assignment of a status value to all individuals in a large society. We have reviewed potential proxies for status in non-human animals, and found that body-size in insects seems to be the only such quantity for which data is currently available for large groups. We present a comparison to our model in section \ref{section:bodysizes}. In the case of humans, socioeconomic data about large groups is available and we justify the use of household income as a proxy for societal status in large human societies and compare this proxy to status distributions from our model in section \ref{section:incomedistributions}.

\subsection{Agonistic interactions in small groups of animals}
\label{section:agonistic}

\paragraph{} In many studies across a wide range of taxa \cite{Shizuka2015}, researchers of animal social behaviour have observed and recorded agonistic interactions between pairs of individuals. These interactions, which include aggressions, physical and non-physical threats, and submissive behaviours \cite{Sapolsky2005} can be considered as ``fights'' in which a winner and loser can be identified. From these observations, an ``interaction matrix'' can be created, with a row and column for each individual, and where each entry $(i,j)$ of the matrix represents the number of times $i$ has defeated $j$ in a pairwise fight. Various methods exist to determine a rank-ordering of the individuals in the society from the data contained in the interaction matrix \cite{Albers2001, DeVries2006, Neumann2011, SanchezTojar2018}. Correlations can then be investigated between hierarchical rank and outcomes such as health, reproductive success, quality of social relationships, and preferential access to resources of the individuals in the social group.

\paragraph{} A common observation in many such studies is that the higher-ranked individual in a pair wins the large majority of the fights against the lower-ranked individual. However, in most datasets, a small number of interactions occur in which the lower-ranked individual wins the fight (called ``reversals'' in the animal behaviour literature). In our model, the distributions of status that most closely resemble this scenario are highly-skewed distributions in which $p \approx 1$ (see Eq \ref{Eq:p}) for many pairs of individuals, such that reversals are rare but still possible. This excludes $\alpha \leq 0$, since reversals are frequent in this range of parameter space, and highlights the relevance of the long-lived states observed in our model (see Fig \ref{fig:stability_diagram}). For small values of $\alpha>0$, the said distributions occur at late stages in the evolution of the system, and for larger values of $\alpha>0$, at the said distributions occur at the stage immediately following the evolution of the system away from the initial condition (see the discussion regarding the three phenomenological stages of the evolution of the system described in section \ref{section:longlived_behaviour} and part E of \nameref{S1_Appendix}).

\begin{figure}[H]
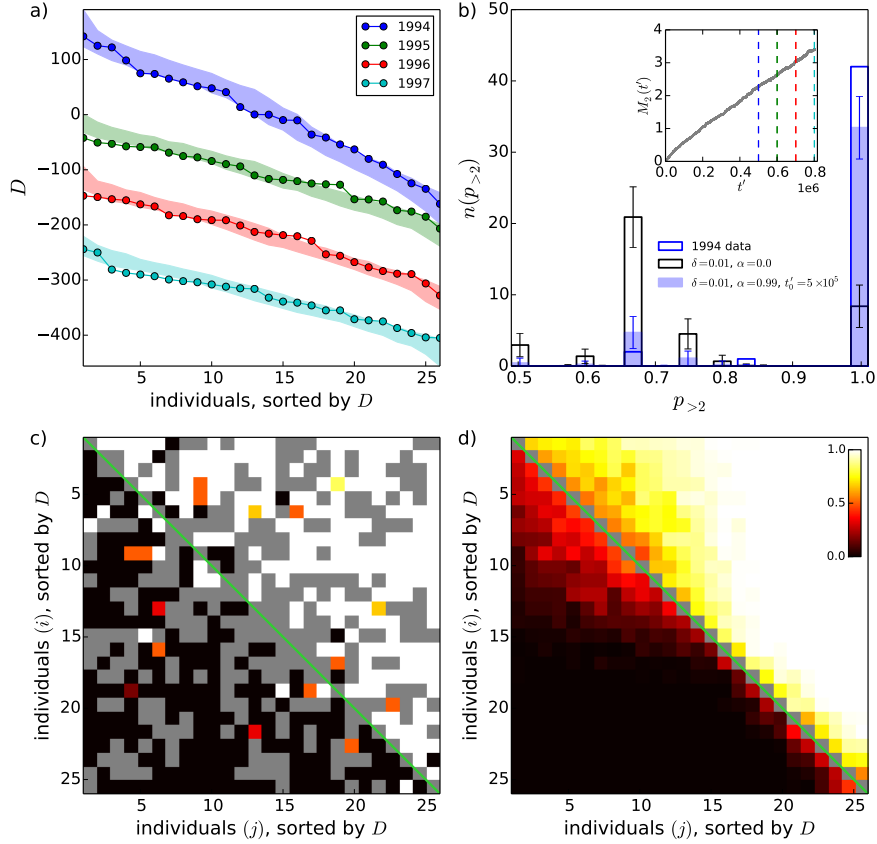

 \centering
 \includegraphics[width=\linewidth]
 {{{D_and_heatmaps_delta0.01_alpha0.99_t0500000_interval100000}}}
 \caption{Comparison of model results with animal interaction data. (a) David's Scores, $D$, calculated from the mountain goat interaction data of Ref. \cite{Cote2000}, for each of four summers from 1994-1997 (solid points). Individuals are ordered by decreasing $D$ along the x-axis, and the value of $D$ is plotted on the y-axis. The coloured bands show $5\%-95\%$ ranges for $D$ calculated from interaction matrices obtained from simulations of the original (two-parameter) model with $\delta=0.01$, $\alpha=0.99$, and $t_0'=5\times10^5$ interactions, where $t_0'=N\;t_0/2$ as per the definition of time in section \ref{section:time}. $n_r=100$ realizations of the simulation were performed. (b) Histogram of $p_{>2}$, the probability that the more successful individual won in a pairwise interaction, considering only those pairs of individuals that engaged in three or more interactions. $p_{>2}$ was calculated from the interaction matrices corresponding to the blue points (animal data from 1994) and blue band (simulation) in panel (a), and for a simulation with $\alpha=0$ (black). For the simulations, the height of each histogram bar shows the average and the error bars show the standard deviation of the number of counts per bin. Inset of (b): $M_2(t)$ for the simulations from which the coloured bands in (a) and filled coloured histogram bars in (b) were obtained, showing (dashed coloured lines) the periods in the simulation during which interaction matrices were recorded. Panels (c) and (d): heatmaps showing individual $i$'s probability of defeating individual $j$ in a pairwise interaction, calculated (c) from the 1994 mountain goat data, and (d) from the simulations from which the blue band in (a) was obtained (averaged over the $n_r=100$ realizations). Grey squares indicate that no interaction occurred between $i$ and $j$. Probability values indicated in the colour-bar in the top-right corner of (d) are applicable to both (c) and (d).}
 \label{fig:Cote_data_comparison}
\end{figure}

\paragraph{} In Fig \ref{fig:Cote_data_comparison}, we compare interaction matrices from an animal observational study with ``simulated interaction matrices'' from our model. We use the largest interaction matrices included in the recent review of Shizuka and McDonald \cite{Shizuka2015}, which were from a study of adult female mountain goats by C{\^{o}}t{\'{e}} \cite{Cote2000}. C{\^{o}}t{\'{e}} published interaction matrices recorded over four summers, from 1994-1997, where $N=26$ individuals were present in all four years. We ran simulations of the model for a system size $N=26$, in which we recorded interaction matrices for the individuals in the simulation. These ``simulated interaction matrices'' were recorded over four separate time periods, where the duration of each time period was equal to the number of interactions recorded during each of C{\^{o}}t{\'{e}}'s summers of observation (C{\^{o}}t{\'{e}} recorded an average of 279 interactions/summer for the 26 goats, using ad libitum sampling). The simulated interaction matrices were each separated by a time period corresponding roughly to the number of pair-wise interactions that $N=26$ female mountain goats are expected to have in one year, estimated here to be approximately $10^5$ interactions from the rate of 3 interactions/individual/hour found in Ref. \cite{Fournier1995}. We allowed each simulation to evolve to a time $t_0 \gg \hat{\tau_1}$ before recording the first simulated interaction matrix, to bypass the initial transient evolution of the simulation away from the egalitarian initial condition. The mountain goat interaction matrices change very little from year to year. This corresponds to a very slowly evolving simulation, where the status distribution remains essentially unchanged between recordings of the simulated interaction matrices, highlighting again the relevance of the long-lived states observed in our model. This occurs for small values of the parameter $\delta$ ($\delta=0.01$ was used in Fig \ref{fig:Cote_data_comparison}).

\paragraph{} In Fig\ref{fig:Cote_data_comparison}a, we show a comparison of the ``David's Score'', $D$, calculated from the simulated and real interaction matrices. $D$ is a commonly-used score that allows a ranking of the individuals in an interaction matrix. It is defined as follows \cite{DeVries2006, David1987}. Let $s_{ij}$ be the number of times individual $i$ has won in an fight against individual $j$, and let $n_{ij}$ be the total number of fights between $i$ and $j$. $P_{ij}=s_{ij}/n_{ij}$ is then the proportion of wins that $i$ has experienced in fights with $j$. The proportion of losses that $i$ has experienced in fights with $j$ is $1-P_{ij} = P_{ji}$, and when $n_{ij}=0$, $P_{ij}$ and $P_{ji}$ are set equal to $0$ \cite{DeVries2006}. Let $w_i=\sum_{j=1,j\neq i}^NP_{ij}$ and $w_{i,2}=\sum_{j=1,j\neq i}^Nw_jP_{ij}$, such that the sum in $w_{i,2}$ is weighted by the $w_j$ of each opponent $j$. Similarly, let $l_i=\sum_{j=1,j\neq i}^NP_{ji}$ and $l_{i,2}=\sum_{j=1,j\neq i}^Nl_jP_{ji}$. The David's Score of an individual is $D_i = w_i+w_{i,2}-l_i-l_{i,2}$. $D$ thus depends not only on the proportions of wins and losses experienced by an individual, but also on the win and loss proportions of those with whom the individual has fought. For example, if an individual $i$ defeats an opponent who has won a large proportion of his or her fights, this causes a large increase $D_i$, whereas if $i$ loses to an individual who has lost a large proportion of his or her fights, this causes a large decrease in $D_i$.

\paragraph{} Fig \ref{fig:Cote_data_comparison}a shows that the set of $D$'s calculated from the simulated interaction matrices resembles the one from the mountain goat interaction matrices, and that this resemblance is maintained after allowing a large number of interactions ($10^5$) to take place between recording the simulated interaction matrices. Fig\ref{fig:Cote_data_comparison}b shows that, in the mountain goats, the more successful individual wins the fight almost all of the time, considering those pairs of individuals that engaged in three or more fights. A similar result was obtained from the simulation for the parameter values used in Fig\ref{fig:Cote_data_comparison}a. Fig \ref{fig:Cote_data_comparison}b also shows that, when $\alpha=0$ (such that $p=0.5$ in Eq \ref{Eq:p}), there are few pairs for which the more successful individual wins all fights (as expected), such that $\alpha>0$ is required in order to have a good comparison with the animal data. Only those pairs of individuals that had engaged in three or more fights were considered in the histograms in Fig \ref{fig:Cote_data_comparison}b, in order to avoid high fluctuations due to small numbers of fights. Also included in Fig \ref{fig:Cote_data_comparison} are heatmap plots showing the probability that individual $i$ defeats individual $j$ in a pairwise fight, calculated from the mountain goat interaction matrices (Fig \ref{fig:Cote_data_comparison}c) and the simulated interaction matrices (Fig \ref{fig:Cote_data_comparison}d), showing visually that reversals are rare but non-negligible.

\paragraph{} Many animal observation studies have found similar interaction matrices to C{\^{o}}t{\'{e}}'s mountain goat matrices, in that they have small numbers of reversals \cite{Shizuka2015}. The results from our model therefore suggest that many animal groups have highly-skewed distributions of status and relatively large values of the authoritarianism, $\alpha$. On the other hand, it is known that animal species with more complex social organization can have a larger number of reversals and intransitive relationships (where individual $i$ wins the majority of fights against $j$, who wins the majority of fights against $k$, who in turn wins the majority of fights against $i$) \cite{Chase2011, Scott1999}. This may correspond to a less highly-skewed distribution of status in our model. Little is presently known about the frequency of reversals and intransitive relationships in large social groups due to a lack of interaction data for large groups. A key problem in this regard is that the proportion of pairs of individuals for which no interactions are observed generally grows with system size in interaction matrices from observational studies, due to the increasing difficulty of observing all pairs \cite{Shizuka2015}.

\paragraph{} Due to the small value of $\delta$ required to obtain quasi-stationarity of the status distribution over large numbers of fights in the simulations in Fig \ref{fig:Cote_data_comparison}, the model suggests that relatively small amounts of status (e.g., one one-hundredth of an individual's current status when $\delta=0.01$) are exchanged in a single interaction. In female mountain goats, rank is strongly correlated with age, and older goats almost always win interactions against younger goats \cite{Cote2000}. Larger values of $\delta$ and smaller values of $\alpha$ may apply in other animal societies in which ranks change more frequently or where age is a less important factor, including in some primate species in which complex power struggles leading to takeovers are commonly observed \cite{Sapolsky1983, DeWaal2007}.

\paragraph{} We also note that for the $N=26$ individuals present in all 4 years of C{\^{o}}t{\'{e}}'s study, there was no tendency for individuals to interact more frequently with those close in rank than with those far away in rank, although such a tendency is observed when considering all individuals in a single summer, as shown in Fig 6 of Ref. \cite{Cote2000} (see part A of \nameref{S2_Appendix} for further details). Our subset of $N=26$ individuals leaves out those individuals who enter the group (primarily, by ageing to the age of maturity of 3 years old) and who leave the group (primarily, by dying at old age) from one year to the next, which indicates that the tendency observed by C{\^{o}}t{\'{e}} for individuals to interact more frequently with those closer in rank is mostly due to interactions involving the oldest and youngest individuals within the group. The absence of this tendency in the $N=26$ subgroup that we consider supports our use of the original (two-parameter) model in the simulations in Fig \ref{fig:Cote_data_comparison}.

\paragraph{} Lastly, for the model simulations shown in Fig \ref{fig:Cote_data_comparison}, the characteristic time $\tau_2 \approx 59$ years, significantly longer than the 12-15 year life expectancy of mountain goats \cite{Guillams2007}. The ultimate collapse of the system to the totalitarian end-state exhibited by our model may therefore not be relevant for mountain goats, since factors such as births, deaths, maturation of juveniles, and immigration, which are not considered in our model but which occur on time-scales significantly shorter than $\tau_2$, will change the long-term dynamics of the system significantly.

\subsection{Proxies for status in large social groups}
\label{section:proxies}

\paragraph{} In the comparison of our model with animal interaction data shown in section \ref{section:agonistic}, we recorded the simulated interaction matrices starting from an arbitrary initial distribution of status, $S(t_0)$. In principle it is possible, for an arbitrary value of $\alpha$, to estimate the $\lbrace S_i \rbrace$ values of the animals in an observed interaction matrix. This would require enough interaction data to estimate the win probabilities for a sufficient number of pairs of individuals such that the ratio $S_i/S_j$ could be calculated from Eq \ref{Eq:p} for these pairs, and then  all $\lbrace S_i \rbrace$ calculated from these ratios. In so doing, one of the $\lbrace S_i \rbrace$ can be set to an arbitrary value since the dynamics of the model are independent of the value of the (conserved) average status, $\bar{S}$, of the system. The data in published animal interaction matrices (including in Ref. \cite{Cote2000}) is insufficient for this purpose, because there are few pairs of individuals where the win probability $p<1$, making it impossible to obtain a meaningful estimate of $\lbrace S_i \rbrace$. Given the difficulty of determining statuses of individuals in the model directly from observed interactions in animal behaviour studies, to compare our model results with real-world societies we seek a measurable quantity that can be used as a proxy for societal status in (large) social groups.

\subsubsection{Intraspecific body size distributions in social insects}
\label{section:bodysizes}

\paragraph{} We have reviewed a number of measurable quantities that may serve as proxies for societal status in non-human animal groups and these are presented in Table S2.2 in \nameref{S2_Appendix}. The only such quantity for which we were able to find data that would allow one to assign a status value to all individuals in a large group ($N \geq 100$) is body size in social insects. Data collected and reviewed by Gouws and co-workers shows that body size distributions in social insects are typically right-skewed as opposed to being normally distributed in non-social insects \cite{Gouws2011}. The distributions of status formed in our model are also right-skewed (e.g. see Fig \ref{fig:PSvS_alpha0}), and thus compare favourably with the body size distributions of social insects. While a more detailed comparison between simulated status distributions and intraspecific body size distributions in social insects would be desirable, it is currently problematic because published data either does not involve individuals from a single colony or does not contain enough information about the distribution to allow a comparison (specifically, both the variance and skewness of the proxy distribution are at least required).

\paragraph{} While we were unable to find other proxy data for societal status in large groups of non-human animals, such data does exist for large groups of humans. We thus focus the remainder of this section on a comparison of the distributions of status produced by our model to a proxy for societal status (household income) in humans.

\subsubsection{Income distributions in humans}
\label{section:incomedistributions}

\paragraph{} In humans, researchers use a theoretical construct called socioeconomic status (SES) to assign positions to individuals in the social hierarchy. Measurable characteristics such as income, education, occupation, and wealth are used as single-factor indicators of SES, or are combined in various ways to obtain composite indicators of SES \cite{Shavers2007}. When income data is available, it is considered to be a critical component in determining SES \cite{Carr2012} and, when used as a single-factor indicator of SES, has been found to have greater explanatory power (for example, of the relationship between SES and mortality) than education or occupation \cite{Duncan2002}. Here, we use household income in Canada and the USA as a proxy for societal status, because income data is readily available in online datasets for these two countries (unlike data on wealth) and it is quantitative (unlike level of education or occupation), making it straightforward to analyze and compare to our model results. We use household income, rather than personal income, as our proxy for societal status, in order to avoid artifacts related to income sharing within a nuclear family unit, such as when a spouse decides not to work or to work below his or her maximum market value \cite{Krieger1997}. We thus follow Ref. \cite{Duncan2002} and many other studies (e.g. \cite{Stewart1992, Soteriades2003, Tandon2012}) and use household income as an established proxy for socioeconomic status.

\paragraph{\textbf{Canadian income distribution}}

\begin{figure}[H]
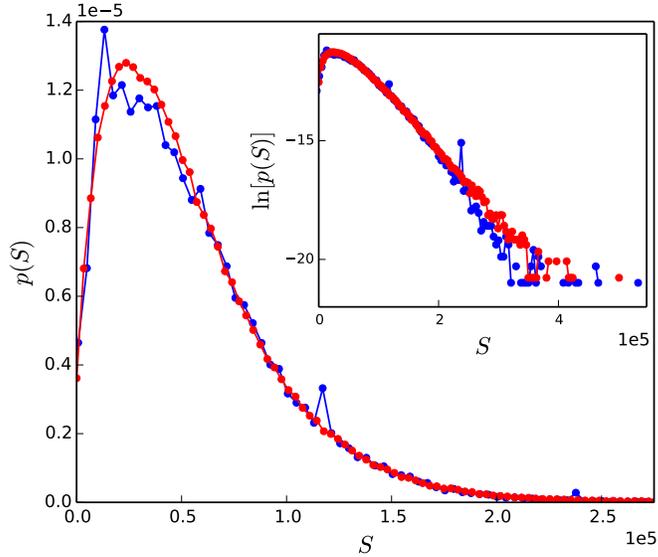

 \centering
 \includegraphics[scale=0.65]
 {{{CAN_fit_orig_model}}}
 \caption{{\bf Fit of original (two-parameter) model status distribution to Canadian household income distribution.} Simulated distribution (red curve, $\delta=0.35$ and $\alpha=0$), compared to the distribution of Canadian household incomes from the year 2000 (blue curve). Scaled variance, $M_2/\bar{S}=0.536$ and skewness, $\gamma=1.3$ for the simulated distribution, and $M_2/\bar{S}=0.534$ and $\gamma=1.4$ for the proxy distribution. The x-axis of the main plot has been cut at $S=2.5 \times 10^5$. The large peak in the data at $\$12,000$ ($S=0.12 \times 10^5$) is most likely due to welfare benefits provided by the government, and the peak at $\$120,000$ ($S=1.2 \times 10^5$) comes from an upper income cutoff applied to the public-use data by Statistics Canada \cite{Statcan2001b}.}
 \label{fig:CAN_inc_distn_onefig}
\end{figure}

\paragraph{} Fig \ref{fig:CAN_inc_distn_onefig} shows the (pre-tax) distribution of Canadian household incomes from the 2001 Canadian census \cite{Statcan2001}, as well as a model-generated status distribution for the two-parameter version of our model (section \ref{section:the_model}). The simulated system contained $N=312,513$ individuals, which is the number of households in the public-use census sample. The average status of each individual was set equal to the average household income in the data, such that $\bar{S}=55,536$. The model parameter $\alpha$ was set equal to 0, in order to produce true steady-state status distributions, for convenience in obtaining a fit to the real data that is independent of observation time, $\tau_{obs}$. The parameter $\delta$ was then adjusted until a good fit was obtained with the income data ($\delta=0.35$ for the plot in Fig \ref{fig:CAN_inc_distn_onefig}). We note that essentially the same status distributions --- albeit not steady-state but only long-lived --- can be obtained for ($\delta, \alpha$) values that trace out an equi-$M_2$ line in $\delta-\alpha$ parameter-space (see Fig \ref{fig:stability_diagram} and part H of \nameref{S1_Appendix}). 

\paragraph{} The original (two-parameter) version of our model is sufficient to fit the Canadian household income distribution. This suggests that our very simple model may capture some essential features of the interactions that give rise to social hierarchy in large groups of individuals. The two model parameters, $\delta$ and $\alpha$, which determine the outcomes of pairwise interactions, may be interpreted as societal features since their values are held constant for all interactions that occur within the society. Particular societies or species may have different values of one or both of these parameters, leading to different societal structures represented by the distribution of status. Additional comments about the potential real-world implications of the time evolution behaviour of the model are included in section \ref{section:discussion}.

\paragraph{} A shortcoming of the Canadian data is that upper income cutoffs were applied to the public-use dataset by Statistics Canada, for the purpose of protecting confidentiality of wealthy individuals. This results in poor data quality at the upper income end of the Canadian household income distribution. However, there is evidence that income distributions from many countries have a low-to-middle-income part that is well described by a distribution function containing an exponential decay, and a separate, high-income part that decays more slowly \cite{Yakovenko2009, Chakrabarti2010}. The inset of Fig \ref{fig:CAN_inc_distn_onefig} shows the Canadian household income distribution and the model-generated status distribution on a log-linear scale, in order to allow inspection of the large-$S$ tail. The expected slower-than-exponential decay of the upper-tail is not seen in this data. This motivates us to examine another dataset in which the distinct low-to-middle and upper income parts of the distribution are present. We do this for the USA household income distribution in the next section.

\paragraph{\textbf{USA income distribution}}

\begin{figure}[H]
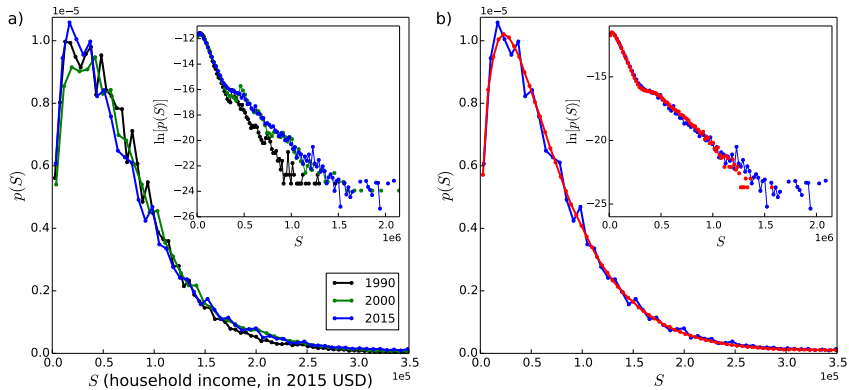

 \centering
 \includegraphics[width=\linewidth]
 {{{USA_multipanel}}}
 \caption{{\bf USA household income distributions and fit of extended model.} (a) USA household income distributions for three different years (indicated in legend). Dollar values have been converted to 2015 values to allow comparison of the different datasets \cite{Ruggles2015b}. A ``break'' in the data separating the high-income tail from the low-to-middle part of the income distribution can be seen in the inset at $S \approx \$250,000$ ($S=0.25 \times 10^6$). Upper-income cutoffs have been applied to the American data by the governmental agency that provided the data, causing the plateau in $\ln[p(S)]$ (inset) for the highest income values. (b) Fit of extended model status distribution to USA household income distribution. Simulated distribution (red curve) with parameters $\delta=0.4$, $\alpha=0$, $\eta=3.5$, $\epsilon=0.08$ compared to the distribution of USA household incomes from the year 2015 (blue curve).}
 \label{fig:USA_data_and_fit}
\end{figure}

\paragraph{} Fig \ref{fig:USA_data_and_fit}a shows United States household income distributions for the years 1990, 2000, and 2015 \cite{Ruggles2015}. The USA datasets are 1-in-100 national random samples of the  population. Dollar values for each of the three datasets have been adjusted to 1999 values in order to permit a comparison across time. In these distributions, the presence of an approximately exponentially-decaying low-to-middle-income part (initial straight-line decrease in the inset of Fig \ref{fig:USA_data_and_fit}a beginning after the peak at $S \approx 0.25 \times 10^5$ and ending at $S \approx 0.25 \times 10^6$), separated from a more slowly-decaying high-income tail is visible. The ``break'' point between the lower and upper parts of the data is present for all three curves and can be seen in the inset of Fig \ref{fig:USA_data_and_fit}a at $S \approx 0.25\times 10^6$. This break is observed in the income distributions of many different countries \cite{Yakovenko2009, Chakrabarti2010}.

\paragraph{} Fig \ref{fig:USA_data_and_fit}b shows the 2015 USA household income distribution and a simulated distribution from the extended model. The main plot (linear scale on the y-axis) shows that the simulated distribution fits well to the low-to-middle income part of the distribution. The inset (logarithmic scale on the y-axis) shows that the simulated distribution also contains a break between the low-to-middle and high-income parts of the distribution, mirroring the break in the data. The value of the parameter $\eta$ was chosen such that $\eta\bar{S} = S_B$, where $S_B$ is the location of the ``break'' point in the data, estimated from the data to be at $\$275,000$. Changes to the value of $\epsilon$ result in a poorer fit (see \nameref{S2_Appendix}, part C). As expected, $\epsilon$ is small, such that only $8\%$ ($\epsilon=0.08$) of the pairwise interactions that would not occur according to the threshold criterion do occur. Two alternative ways to restrict the pairwise interactions between competing individuals that lead to quantitatively similar behaviour are described in part E of \nameref{S2_Appendix}; for example, a simple, additional extension to the model that allows all individuals with statuses greater than $\eta\bar{S}$ to interact with each other shows an improved fit to the proxy data.

\begin{figure}[H]
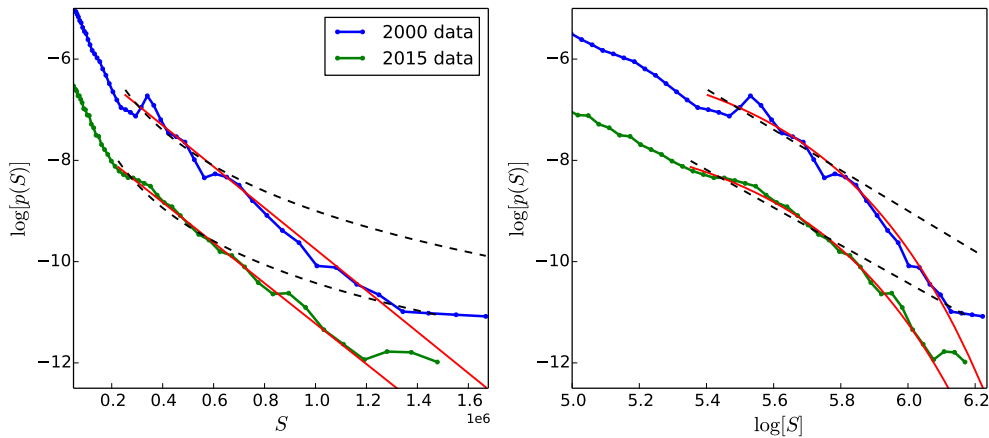

 \centering
 \includegraphics[width=\linewidth]
 {{{PDF_EXP_PL_twopanel_fulltail}}}
 \caption{{\bf Functional form of high-income tail of USA household income distribution.} Power-law (dashed black line) and exponential (solid red line) distributions with lower and upper bounds on the fitted distribution chosen to correspond to the full high-income tail. $S$ represents USA household income data in 1999 USD. The curves for 2015 have been shifted down in the plots for better visualization.}
 \label{fig:PL_EXP_fit_real_data_fulltail}
\end{figure}

\paragraph{} Several econophysics studies have found power-law distributions in the high-income tail of personal income distributions \cite{Yakovenko2009, Chakrabarti2010}. However, a graphical analysis (Fig \ref{fig:PL_EXP_fit_real_data_fulltail}) comparing best fits of exponential and power-law distributions to the high-income tails of the 2000 and 2015 USA household income distributions shows that the household income data that we use as a proxy for societal status is not consistent with a power-law distribution but is consistent with an exponential distribution over large ranges. This is also confirmed by Kolmogorov-Smirnov tests for both distribution types. Further details are provided in \nameref{S2_Appendix} (part D) \cite{Baro2012, SAS2000, Wang2001, Capasso2009}.

\paragraph{} The distinct low-to-middle and upper status parts of the data show the presence of two distinct groups or classes within the society. In the model, these two distinct groups emerge when the frequency of interactions between individuals with large differences in status is reduced. The model thus suggests that societal conditions that limit interactions between individuals with large differences in societal status may produce or maintain distinct social classes. Individuals belonging to the upper status class may have a self-interest in reinforcing such societal conditions in order to preserve their positions in society. Societies with policies that promote interactions between individuals with large differences in societal status may have less distinct class structures.

\section{Discussion} 
\label{section:discussion}

\paragraph{} We have presented a simple winner-loser model of the formation and evolution of social hierarchy, based solely on interactions (fights) between individuals that result in the transfer of societal status from the loser of the fight to the winner. We showed that the model exhibits regions in parameter-space in which the asymptotic distributions of status produced by the model either show a continuous unimodal behavior or take on a degenerate form, in which a single individual possesses all of the society's status. In the latter case, intermediary distributions are long-lived for small positive values of the model parameters. Here, ``long-lived'' refers to quasi-stationarity of the status distribution, where, over a sufficiently short observation time, the status distribution remains essentially unchanged. This is quantified through the characteristic time $\tau_2$, which controls the evolution of the system toward the end-state.

\paragraph{} Our model thus suggests that there are two fundamental characteristics of status-determining interactions in a society --- the level of intensity of interactions ($\delta$) and the degree of authoritarianism ($\alpha$) --- that determine both the outcomes of the interactions and whether the society's structure will be stable or preserved for long times before undergoing eventual deterioration. These two parameters together with optional parameters restricting the interactions between individuals control the shape of the (intermediary) status distribution, which becomes more unequal (larger variance) as either $\alpha$ or $\delta$ is increased.

\paragraph{} In comparing the status distributions produced by simulations of the original (two-parameter) and extended models with the proxies for societal status in section \ref{section:proxies}, the parameter $\alpha$ was set equal to $0$. However, as shown in \nameref{S1_Appendix} (part H), essentially the same long-lived status distributions can be obtained for ($\delta, \alpha$) values that trace out an equi-$M_2$ line in $\delta-\alpha$ parameter-space. Starting with a given value of $\delta$ and $\alpha=0$, one can locate an equi-$M_2$ line in Fig \ref{fig:stability_diagram}. Then, following the equi-$M_2$ line in the direction of increasing $\alpha$, one eventually arrives at the transition between long-lived states and runaway (where the location of the transition depends on the time over which the society is observed). If the value of $\delta$ of a society can be determined by fitting the model (with $\alpha=0$) to a proxy for the society's status distribution, then the corresponding equi-$M_2$ line may provide an indication of the maximum level of authoritarianism for which essentially the same status distribution can be maintained over a specific time interval. This maximum level of authoritarianism would be reached when the equi-$M_2$ line intersects with the boundary separating regions II and III for this specific time interval, see Fig \ref{fig:stability_diagram} for examples.

\paragraph{} Similarly, the presence of a characteristic time scale controlling the longevity of the intermediary distributions suggests a limit on the extent to which societal inequality can increase (e.g., due to societal changes that cause an increase of one or more of the parameters) before a runaway deterioration occurs.  Whether a real society in fact approaches the end-state might depend on how this characteristic time scale compares with other time scales neglected in our model, such as the rate at which the society experiences external perturbations including wars with other societies or major environmental changes, and internal perturbations related to the effects of birth, death, immigration, and aging of individuals. For example, in the comparison of the model with agonistic interactions in animals shown in section \ref{section:agonistic}, $\tau_2 \approx 59$ years, significantly longer than the 12-15 year life expectancy of mountain goats \cite{Guillams2007}, such that births, deaths, maturation of juveniles, and immigration may change the long-term dynamics of the system in such a way that it remains far from the end-state predicted by our model.

\paragraph{} Extending our original (two-parameter) model by introducing an additional model rule that adjusts the probability with which individuals interact with one another based on the differences in their statuses, we can produce stable and long-lived status distributions which have identifiable low-to-middle and upper status regions. The status distributions in our extended model show good agreement with the distribution of household incomes in the USA (see Fig \ref{fig:USA_data_and_fit}b), which we use as a proxy for societal status in large social groups. This appears to be the first model in which the two-part structure of the proxy distribution emerges by self-organization based solely on interacting individuals, without requiring any exogenous influence such as redistribution through taxation. Several analyses of personal income distributions have found that the low-to-middle-income part of the distribution decays exponentially and that the high-income tail decays like a power-law \cite{Yakovenko2009, Chakrabarti2010}. Based on these analyses, various econophysics models have been propopsed with the goal of generating distributions of income with two-part shapes in which the lower-to-middle part of the income distribution follows an exponential distribution, and the upper tail follows a power-law distribution \cite{Slanina2013, Richmond2013, Boghosian2016b}. Among the models that generate a distribution of income or wealth as a self-organizing process based on interactions between individuals, several are able to produce either a power-law decay in the upper tail, or an exponential decay in the lower part of the distribution, but not both. The status distributions produced by the original (two-parameter) version of our model do not have two-part structures, and therefore never contain a ``break" in the large-$S$ tail similar to that seen in the insets of Fig \ref{fig:USA_data_and_fit}. But the extended version of our model can produce two-part structures, where both the regime leading up to and the regime following the ``break" decay exponentially. Our model thus suggests that societal structures containing distinct social classes arise when interactions between individuals with large differences in social status are limited or restricted, as occurs, for example, in residential segregation in the USA \cite{Massey1993, Iceland2006, Bischoff2014}.

\paragraph{} A necessary foundation for more advanced studies of social hierarchy is the exploration of the simplest possible realistic models, including the determination of their limits. This was the goal of the present article. Future, network-oriented models may incorporate features such as the histories of interactions between individuals \cite{Davidsen2002, Ebel2003, PinterWollman2014} and cooperative behaviours including the formation of coalitions \cite{Flinn2012, Wilson2003, Sapolsky1992, Hemelrijk2017} and mobbing \cite{Leymann1996, Graw2007}. Such models may provide deeper understanding about the origins and evolution of societal structures, including the mechanisms responsible for societal destabilization or collapse.

\newpage
\setcounter{section}{0}
\setcounter{figure}{0}
\renewcommand{\thesubsection}{S1.\Alph{subsection}}
\renewcommand{\figurename}{Fig}
\renewcommand\thefigure{S1.\arabic{figure}}

\section*{Appendix S1} \label{S1_Appendix}
\addcontentsline{toc}{section}{Appendix S1}

\subsection{Comparison of simulation results to Ispolatov et al. analytic solution}
\label{section:SI_time_comparison}

\begin{figure}[H]
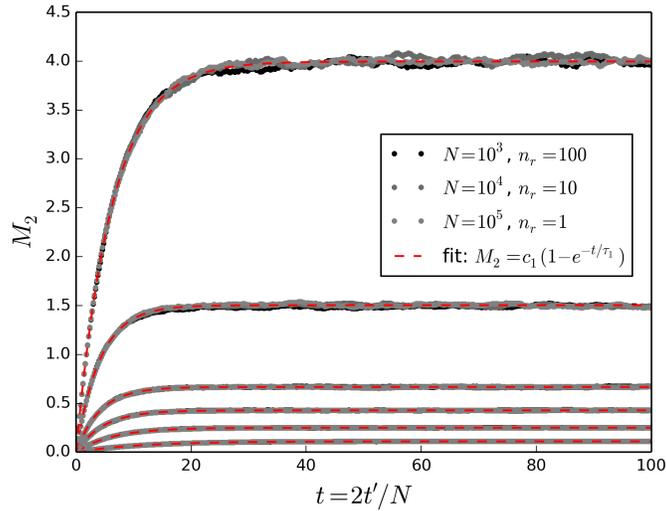

 \centering
 \includegraphics[scale=0.6]{{{M2_v_t_alpha0_vary_delta_vary_N}}}
 \caption{Evolution of $M_2(t)$ for $\alpha=0$ and various values of $\delta$ (from bottom, $\delta=$ 0.1, 0.2, 0.3, 0.4, 0.6, 0.8). Time is scaled so that $t=2t'/N$. Fits are for $N=10^5$. Data averaged over $n_r$ realizations of the simulation.}
 \label{fig:M2_v_t_alpha0_varyN}
\end{figure}

\paragraph{} In order to demonstrate that the definition of time as $t=2t'/N$ in the simulation corresponds to the definition of time in the analytical solution of Eq 2 of the main text, we show (Fig \ref{fig:M2_v_t_alpha0_varyN}) a plot of $M_2(t)$ when the parameter $\alpha=0$ with time defined as $t=2t'/N$. Each cluster of curves in the figure corresponds to a particular value of the model parameter $\delta$. Within a single cluster of curves, simulation results are shown for three different system sizes, $N$. The red line shows a fit of Eq 4, where the fit parameters $\tau_1$ and $c_1$ are equal to their corresponding analytic values from Eq 2. For example, for $\delta=0.8$ and $\bar{S}=1$, the fit parameters are $c_1=4.00$ and $\tau_1=6.29$, whereas Eq 2 gives $\hat{c_1}=4$ and $\hat{\tau_1}=6.25$.

\newpage
\subsection{Time evolution of distributions}

\subsubsection{Beginning from an egalitarian initial condition}

\begin{figure}[H]
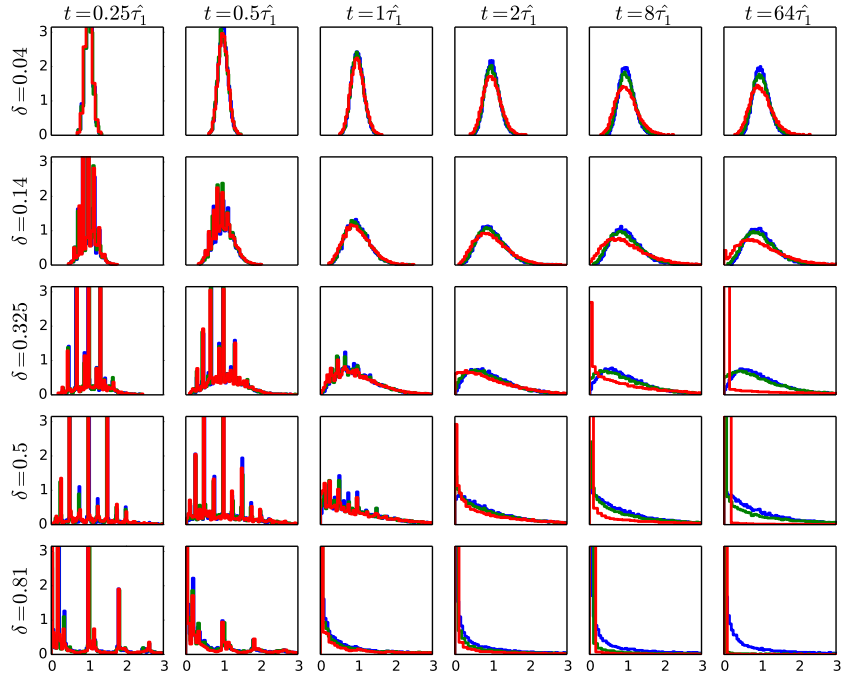

 \centering
 \includegraphics[width=\linewidth]
 {{{time_evol_grid_alpha_0.5and0.2and0.0_icEgal_N1000_nr20}}}
 \caption{Time evolution of distributions for different values of $\delta$, with $\alpha=0$ (blue), $\alpha=0.2$ (green), and $\alpha=0.5$ (red), starting from an initial condition in which all individuals have status $S=1$ (``egalitarian initial condition"). x-axes in the sub-plots correspond to status, $S$, and y-axes to the probability density, $p(S)$. $N=10^3$, $n_r=20$.}
 \label{fig:time_evol_icEgal}
\end{figure}

\paragraph{} Fig \ref{fig:time_evol_icEgal} shows the time evolution of systems prepared in an ``egalitarian" initial condition, in which all individuals have the same initial status of $S=1$. Each row in the figure corresponds to a different value of $\delta$, and each column to an instant in time, with time increasing from left to right across the subplots.  For all values of $\delta$, the system undergoes an early-time evolution away from the initial condition. When $\alpha=0$ (blue curves), the system arrives at a steady-state in which the distribution of status remains constant in time. The steady-state distribution is obtained within a time equal to a few multiples of $\hat{\tau_1}$, as expected from Eq 4 of the main text. When $\alpha>0$ (green and red curves), the distribution continues to evolve after the transient period, and approaches the totalitarian end-state at a rate that depends on $\delta$ and $\alpha$.

\subsubsection{Beginning from a uniform initial condition}
\label{section:SI_time_evol_uniform_distn}

\paragraph{} Figures showing the evolution of the model beginning from a uniform initial condition are included below, for the same values of $\delta$ and $\alpha$ shown in Fig. \ref{fig:time_evol_icEgal}. In the uniform initial condition, the initial status of each individual is chosen randomly from a uniform distribution with average status $\bar{S}=1$. The time evolution beginning from this initial condition is qualitatively the same as for the egalitarian initial condition.

\begin{figure}[H]
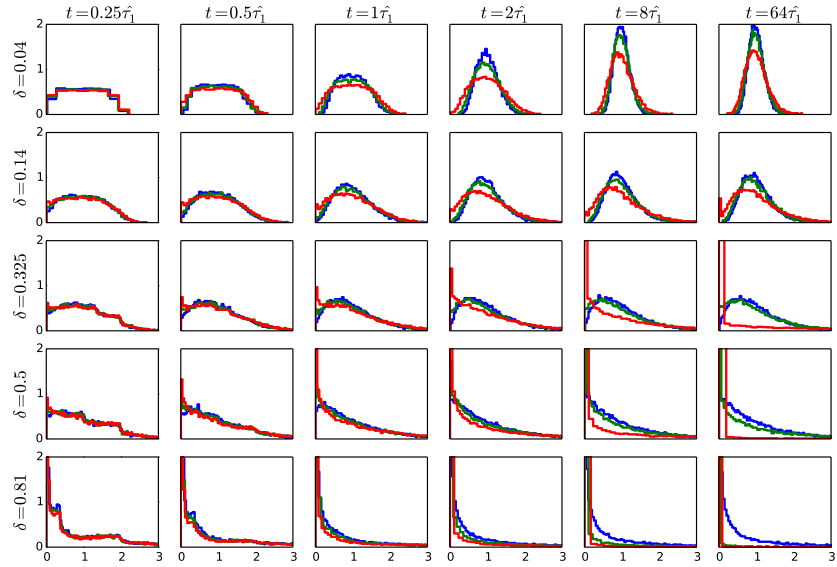

 \centering
 \includegraphics[width=\linewidth]
 {{{time_evol_grid_alpha_0.5and0.2and0.0_icUnif_N1000_nr20}}}
 \caption{Time evolution of distributions for different values of $\delta$, with $\alpha=0$ (blue), $\alpha=0.2$ (green), and $\alpha=0.5$ (red), starting from an initial condition in which status is uniformly distributed (``uniform initial condition"). x-axes in the sub-plots correspond to status, $S$, and y-axes to the probability density, $p(S)$. $N=10^3$, $n_r=20$.}
 \label{fig:time_evol_icUnif}
\end{figure}

\newpage
\subsection{Behaviour of $\tau_1$ and $\tau_2$ as a function of system size, $N$}
\label{section:SI_chartimes_v_N}

\begin{figure}[H]
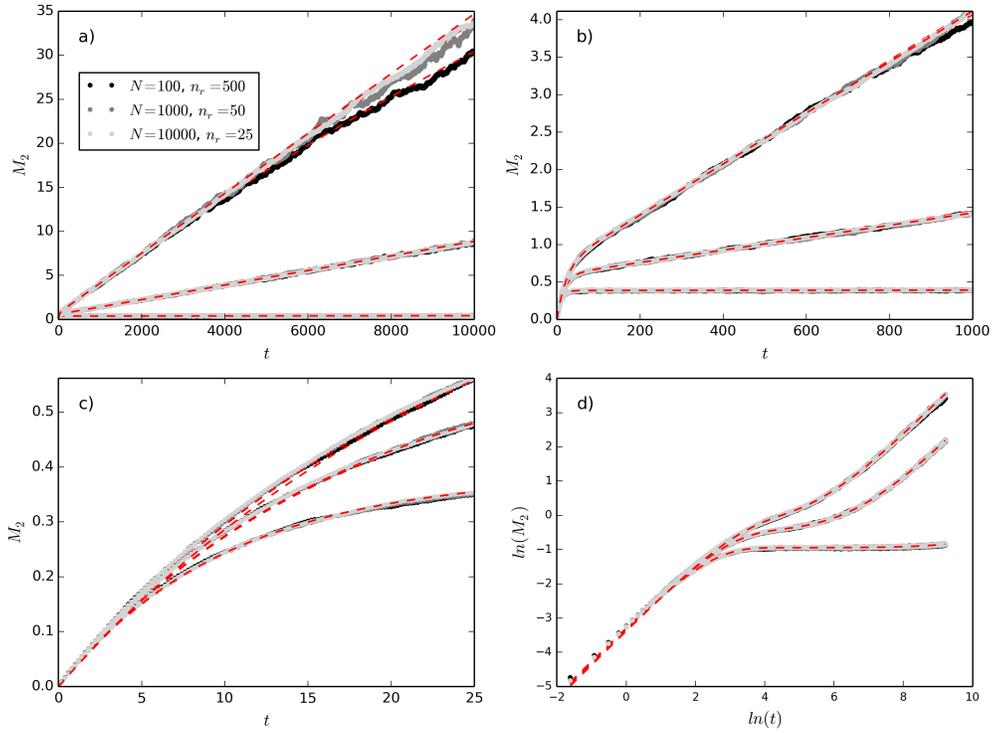

 \centering
 \includegraphics[width=\linewidth]
 {{{tau1_tau2_vary_N}}}
 \caption{Evolution of $M_2(t)$ for $\delta=0.2$ and $\alpha>0$ (from bottom, $\alpha=$ 0.3, 0.5, 0.6) for different system sizes $N$. (a)-(c) different (linear) scales on x and y axes. (d) log-log scale. Fits are of Eq 6 (red dashed lines) to simulated data.}
 \label{fig:tau1_tau2_vary_N}
\end{figure}

\paragraph{} In order to examine the behaviours of the characteristic times $\tau_1$ and $\tau_2$ as a function of system size, $N$, plots of $M_2(t)$ for different system sizes are shown in Fig \ref{fig:tau1_tau2_vary_N}. The parameters controlling the early-time behaviour of $M_2$ ($c_1$ and $\tau_1$) remain constant as system size, $N$, is increased. The parameters controlling the long-time behaviour of $M_2$ increase linearly with $N$. For example, for $\alpha=0.5$, the values of $\tau_2$ extracted from a fit of Eq 6 are $1.17 \times 10^{5}$, $1.18 \times 10^{6}$, and $1.21 \times 10^{7}$ for system sizes $N = 10^{2}$, $10^{3}$, and $10^{4}$, respectively. Likewise, the parameter $c_2$, which represents the ``upper plateau" end-state value of $M_2$ increases linearly with $N$ in the large-$N$ limit (Eq 5).

\newpage
\subsection{Proof that $N^2$ fights are required to reach end-state when $\delta=1$ and $\alpha=\infty$}
\label{section:SI_proof_configurationa_reasons}

\paragraph{} In order to explore how the long-time behaviour of the evolution of the status distribution depends on system size $N$, we consider the extreme scenario of a system in which the higher-status individual in each fight wins with probability $p=1$, and in which the winner of each fight receives all of the loser's status, leaving the loser with zero status. This extreme scenario corresponds to $\delta=1$ and $\alpha=\infty$. In the model presented in the main text, $\delta$ is strictly less than $1$, meaning that no single individual can ever possess all of the societal status of the system. However, in the following we set $\delta=1$ in order to allow the system to obtain (after a finite time and for a finite system size) the end-state in which a single individual possesses all of the societal status and all other individuals have zero status. This simplifies the analysis, leading to the proof described below.

\paragraph{} The goal of this exercise is to determine the time, $\tau_{end}$, required to reach the totalitarian end-state beginning from an egalitarian initial condition. To do so, we will work backwards from the totalitarian end-state, in which a single individual possesses all of the societal status.

\paragraph{} Working backwards from the end-state, we can consider the sequence of events that were required to arrive at the end-state. The first required event, working backwards, is a fight between two individuals possessing non-zero status. Since $\delta=1$, one of these individuals loses and exits the fight with status equal to zero. The other individual wins the fight and receives all of the status of the loser, which, added to its pre-fight status, equals the total status of the system. Therefore, the end-state is preceded by a configuration of the system in which only two individuals possess non-zero status. The probability that these two particular individuals are selected to fight each other is the probability of selecting the first individual OR the second individual AND the probability of selecting the remaining individual:

\begin{equation}
\rho_2 = \left(\frac{1}{N}+\frac{1}{N}\right)\frac{1}{N-1} = \frac{2}{N(N-1)}.
\end{equation}

\paragraph{} Continuing to work backwards, in order to obtain the configuration of the system in which there are two individuals with non-zero status, a preceding configuration having three individuals with non-zero status is required. The probability that two of these three individuals are selected to fight is: 

\begin{equation}
\rho_3 = \binom 32\frac{2}{N(N-1)}.
\end{equation}

Following through with this logic, we find $\rho_N = \binom N2 \frac{2}{N(N-1)}$. Now, the expected number of fights that will elapse between the required configuration $k$ and the subsequent required configuration $k-1$ in this sequence is $\rho_k^{-1}$.  Therefore, the expected number of fights to reach the end-state is: 

\begin{eqnarray}
\label{Eq:tau_end_prime}
\tau_{end}' & = &  \frac{N(N-1)}{2}\left(\frac{1}{\binom 22}+\frac{1}{\binom 32}+\frac{1}{\binom 42}+\ldots + \frac{1}{\binom N2} \right) \nonumber \\
& = & N(N-1) \sum_{k=2}^N \frac{1}{k(k-1)}.
\end{eqnarray}

\paragraph{} The sum in Eq. \ref{Eq:tau_end_prime} approaches 1 for large $N$, such that $\tau_{end} \approx N^2$. The time to reach the end-state when $\delta=1$ and $\alpha=\infty$ is therefore $\tau_{end} \approx 2\tau_{end}'/N = 2N$, for large $N$. 

\paragraph{} This proof is included as a demonstration of the configurational reasons why the long-time (approach to the end-state) evolution of the model dynamics increases in proportion to the system size $N$.

\newpage
\subsection{Phenomenology of the time-evolution of status distributions when $\alpha>0$}
\label{section:SI_phenom}

\paragraph{} In this section we provide some additional details about the phenomenology of the evolution of the status distributions when $\alpha>0$. 

\begin{figure}[H]
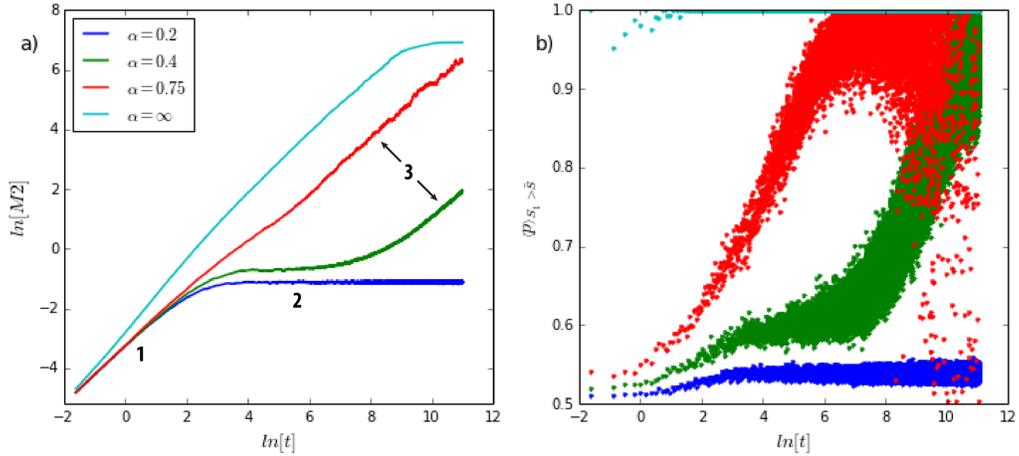

 \centering
 \includegraphics[width=\linewidth]
 {{{phenomenology_delta0.2vary_alpha_N1000_edited2}}}
 \caption{(a) $M_2(t)$ on a log-log scale, with four stages of evolution indicated. $\delta=0.2$ for all curves. (b) Probability $p$ (averaged over preceding 100 fights, considering only those fights in which $S_1>\bar{S}$) as a function of time. $N=1000$ (a and b) and $n_r=10$ (a) and $n_r=1$ (b)}
 \label{fig:phenom_M2}
\end{figure}

\paragraph{} Fig \ref{fig:phenom_M2}a shows a plot of $M_2(t)$ for a system with $N=1000$, $\delta=0.2$, and several values of $\alpha$. 

\begin{figure}[H]
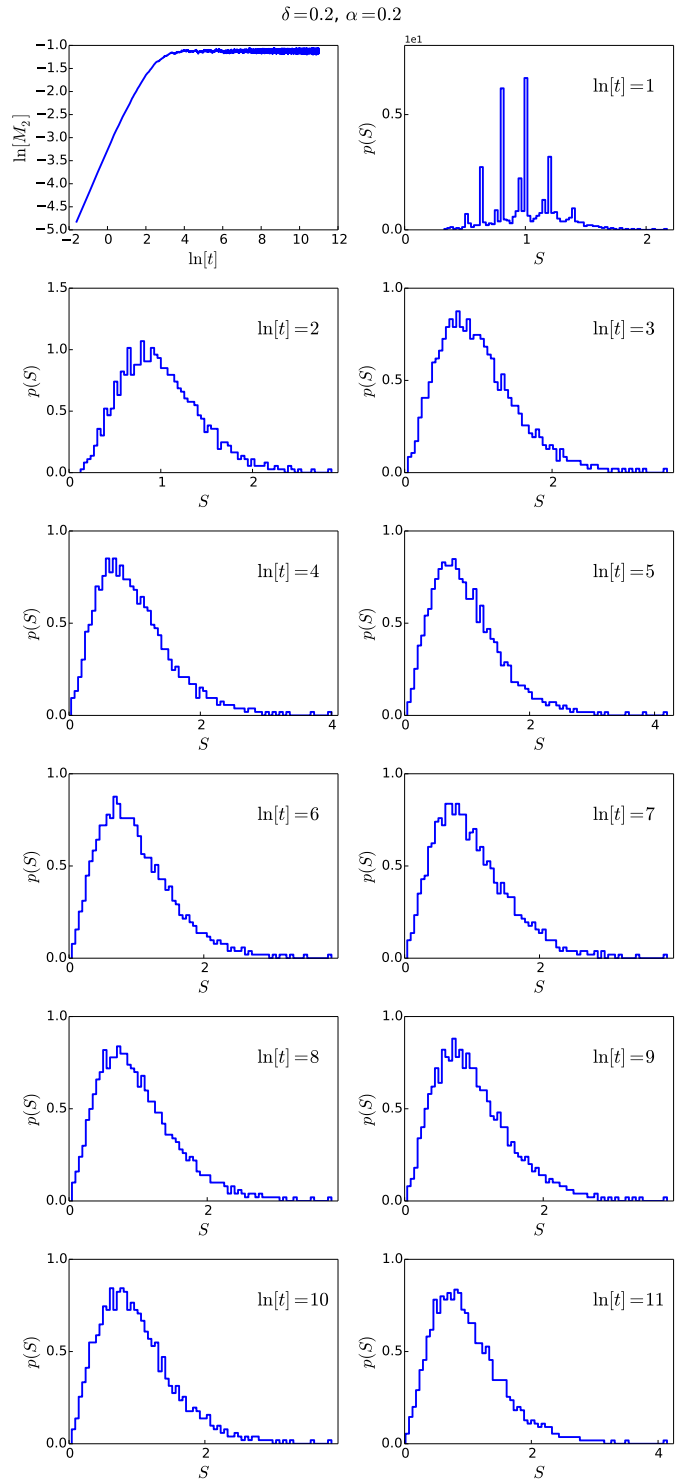

 \centering
 \includegraphics[height=\textheight]
 {{{phenom_hists_v_time_linlin_delta0.2_alpha0.2}}}
 \caption{Time evolution of distributions for $\delta=0.2$, $\alpha=0.2$. $N=1000$, $n_r=10$.}
 \label{fig:phenom_hists_v_time_linlin_alpha0.2}
\end{figure}

\begin{figure}[H]
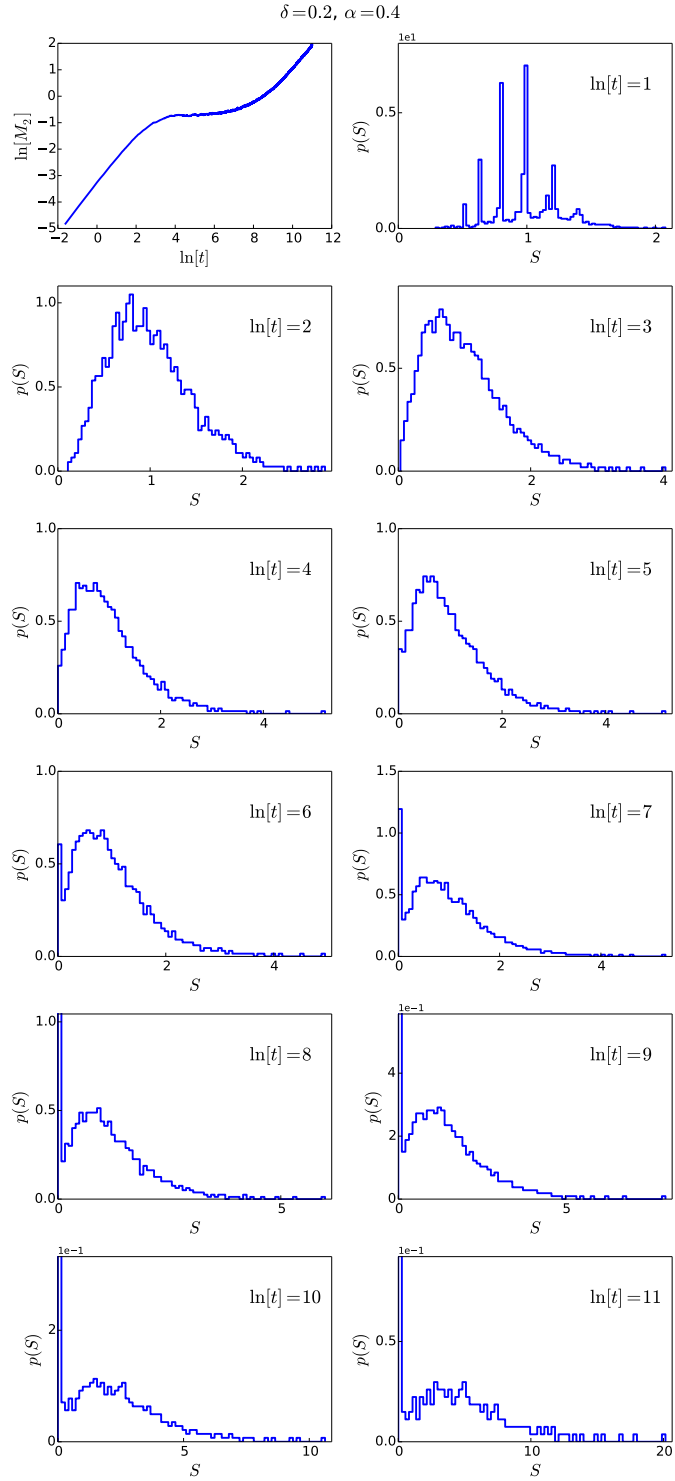

 \centering
 \includegraphics[height=\textheight]
 {{{phenom_hists_v_time_linlin_delta0.2_alpha0.4}}}
 \caption{Time evolution of distributions for $\delta=0.2$, $\alpha=0.4$. y-axis scale adjusted in plots for $\ln[t]=8$ to $11$ to allow visualization of large-$S$ range of distribution. $N=1000$, $n_r=10$.}
 \label{fig:phenom_hists_v_time_linlin_alpha0.4}
\end{figure}

\paragraph{} Three distinct stages in the evolution of $M_2(t)$ can be identified, as indicated on the log-log plot of $M_2(t)$ shown in \ref{fig:phenom_M2}a. First (stage 1), there is a rapid change away from the initial condition into a distribution similar in shape to the $\alpha=0$ (true steady-state) distribution. This distribution shape is essentially maintained during stage 2, such that $M_2(t)$ changes only very slowly with time. For example the curve corresponding to $\alpha=0.2$ (blue) in Fig \ref{fig:phenom_M2}a remains in stage 2 over the time scale of the simulation (to demonstrate this, distributions corresponding to specific points in time along this curve are shown in Fig \ref{fig:phenom_hists_v_time_linlin_alpha0.2}). The duration of stage 2 decreases with increasing $\alpha$, as can be seen in the curve corresponding to $\alpha=0.4$ (green) in Fig \ref{fig:phenom_M2}a (distributions corresponding to specific points in time along this curve are shown in Fig \ref{fig:phenom_hists_v_time_linlin_alpha0.4}). When $\alpha$ is increased further, stage 2 essentially disappears (red curve in Fig \ref{fig:phenom_M2}a -- distributions corresponding to specific points along this curve are shown in Fig \ref{fig:phenom_hists_v_time_linlin_alpha0.75}).

\begin{figure}[H]
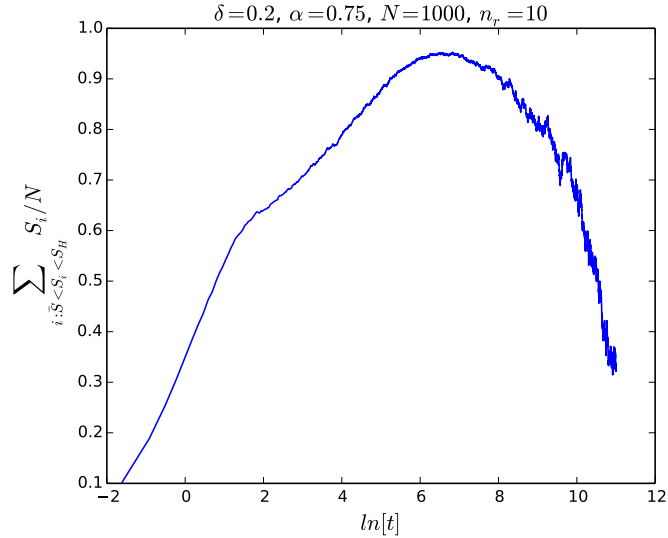

 \centering
 \includegraphics[scale=0.6]
 {{{Sbourg_delta0.2_alpha0.75_N1000_nr10}}}
 \caption{Sum of statuses of all individuals with $\bar{S} < S < S_H$ vs. $t$, for $\delta=0.2$ and $\alpha=0.75$. $N=1000$, $n_r=10$.}
 \label{fig:phenom_Sbourg}
\end{figure}

\begin{figure}[H]
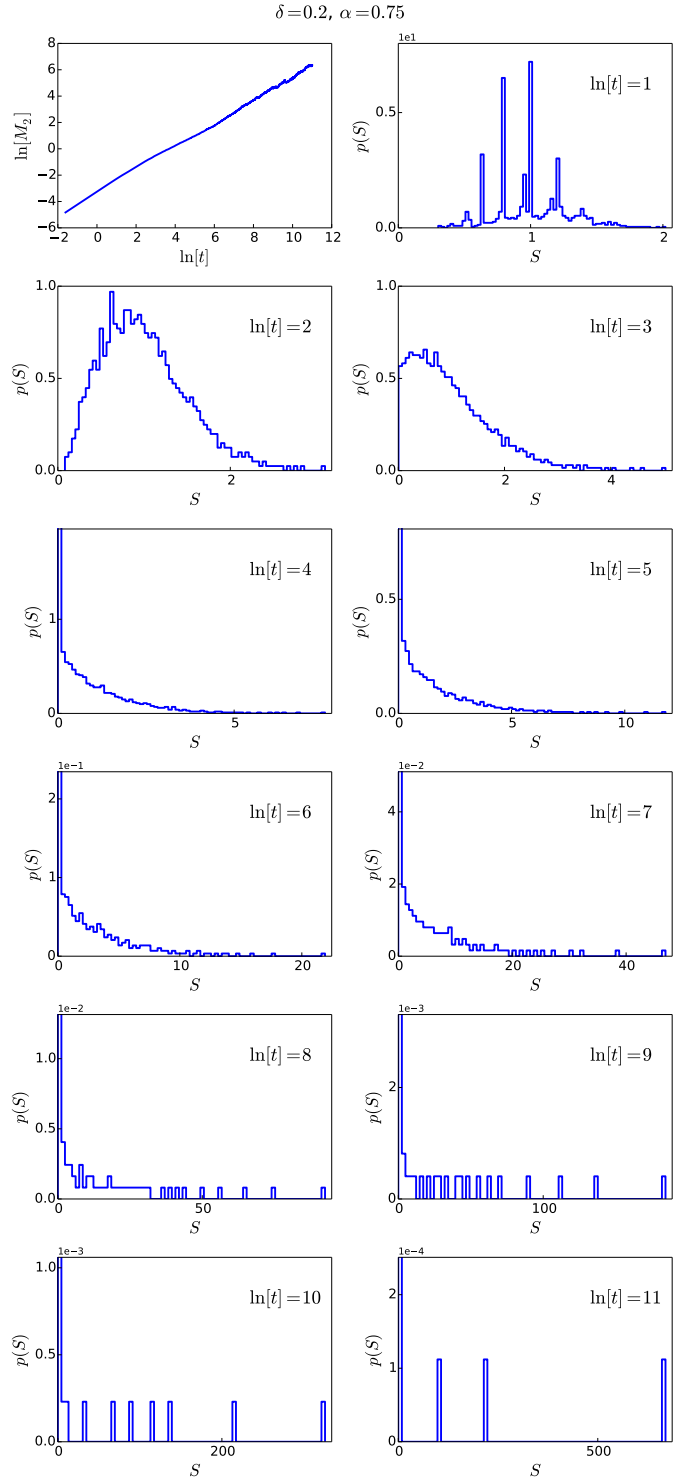

 \centering
 \includegraphics[height=\textheight]
 {{{phenom_hists_v_time_linlin_delta0.2_alpha0.75}}}
 \caption{Time evolution of distributions for $\delta=0.2$, $\alpha=0.75$. y-axis scale adjusted in plots for $\ln[t]=4$ to $11$ to allow visualization of large-$S$ range of distribution. $N=1000$, $n_r=10$.}
 \label{fig:phenom_hists_v_time_linlin_alpha0.75}
\end{figure}

\paragraph{} Transitioning from stage 2 to stage 3, the distribution accumulates many individuals with very low statuses, and several higher-status individuals emerge. Stage 3 then consists of the slow evolution of this highly skewed distribution. A particular characteristic of stage 3 is that, for fights involving at least one individual who still retains sizeable status (e.g., $S_1>\bar{S}$), the probability $p \to 1$. This is illustrated in Fig \ref{fig:phenom_M2}b, in which $p$ is averaged over the preceding 100 fights, considering only those fights in which $S_1 > \bar{S}$, giving the quantity $\left< p \right>_{S_1>\bar{S}}$. The fact that $\left< p \right>_{S_1>\bar{S}} \to 1$ signifies that the higher-status individual in any given fight (excluding fights in which neither individual has greater than average status) is virtually guaranteed to win. For finite values of $\alpha > 0$, in order for this to occur, there must be a large relative status between the two individuals. Stage 3 is therefore characterized by a distribution in which the ratio of $S_2/S_1$ for any two randomly-selected individuals is small enough such that $\left< p \right>_{S_1>\bar{S}} \to 1$  for fixed $\alpha$ (as per Eq 1). The large-time decrease in $\left< p \right>_{S_1>\bar{S}}$ (e.g. for the red curve corresponding to $\alpha=0.75$ in Fig \ref{fig:phenom_M2}b, at $\ln[t]=8$ to $\ln[t]=11$) occurs as individuals with $\bar{S}<S<S_H$ begin to experience decreases in status. Here, $S_H$ is the status of the highest-status individual. This is illustrated in Fig \ref{fig:phenom_Sbourg}, which plots the sum of statuses of all individuals with $\bar{S} < S < S_H$ as a function of time.

\paragraph{} The transition from stage 3 to the totalitarian end-state involves the decrease in the sum of statuses shown in Fig \ref{fig:phenom_Sbourg} following the maximum at $\ln[t] \approx 7$. This shows that in its evolution toward the end-state, the distribution must pass through a series of configurations (stage 3) in which there are a number of individuals with higher-than-average statuses, these individuals having extracted their large statuses from the mass of very low status individuals that accumulates in the transition from stage 2 to stage 3. Only after stage 3 has been obtained are the conditions present in which fights between high status individuals can lead to the emergence, in the end-state, of a single individual with status approaching the total status of the system.

\begin{figure}[H]
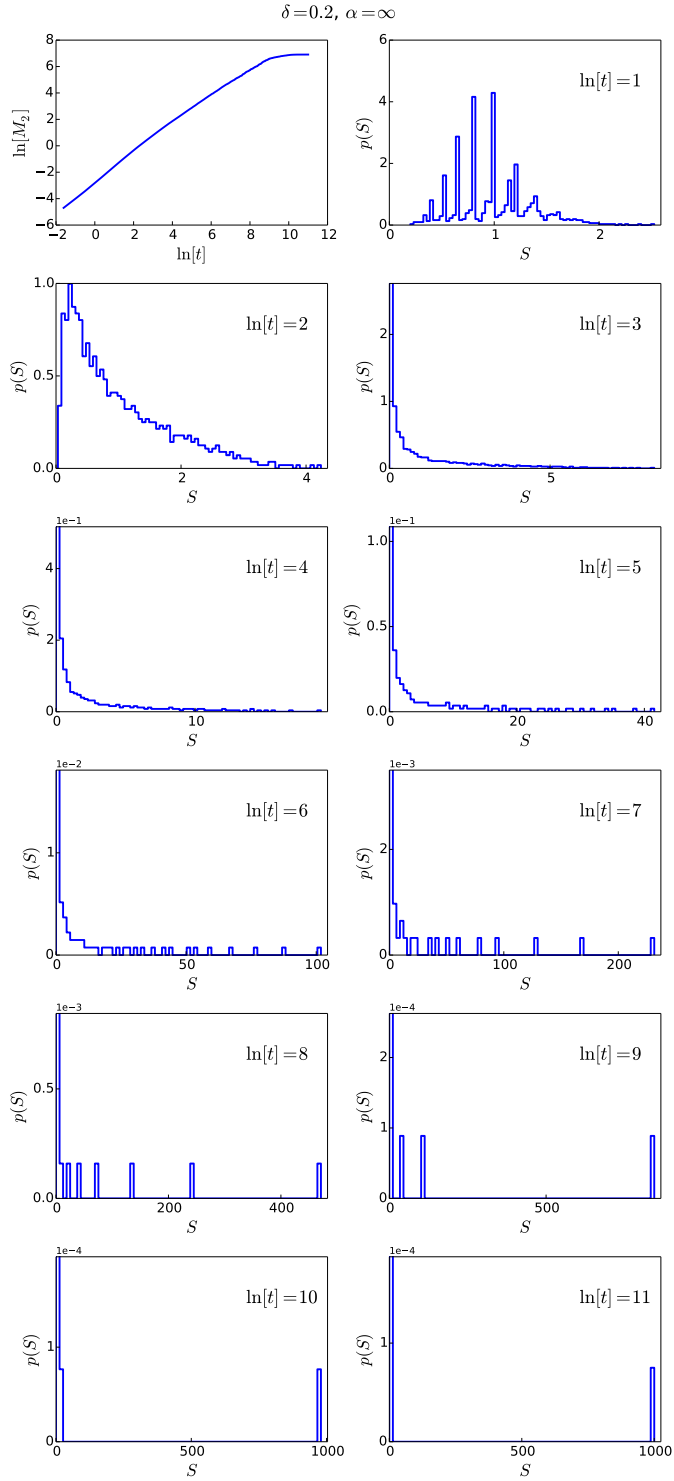

 \centering
 \includegraphics[height=\textheight]
 {{{phenom_hists_v_time_linlin_delta0.2_alphaInf}}}
 \caption{Time evolution of distributions for $\delta=0.2$, $\alpha=\infty$. y-axis scale adjusted in plots for $\ln[t]=3$ to $11$ to allow visualization of large-$S$ range of distribution. $N=1000$, $n_r=10$.}
 \label{fig:phenom_hists_v_time_linlin_alphaInf}
\end{figure}

\paragraph{} In the end-state, $S_H \to N\bar{S}$ and $M_2(t) \to c_2$.  The fourth (light blue) curve in Fig \ref{fig:phenom_M2} corresponds to $\alpha = \infty$, signifying that the higher-status individual in any fight is guaranteed to win (several points in Fig \ref{fig:phenom_M2}b for which $\left< p \right>_{S_1>\bar{S}} < 1$ for this curve result from fights in which $S_1=S_2$ due to the egalitarian initial condition). Distributions at specific points along this curve are shown in Fig \ref{fig:phenom_hists_v_time_linlin_alphaInf}.

\paragraph{} A plot showing the same quantities as in Fig \ref{fig:phenom_M2}, but where $\alpha$ is fixed and $\delta$ is varied, is included below for comparison. As can be seen, as $\delta$ is increased, the duration time of stage 2 decreases.

\begin{figure}[H]
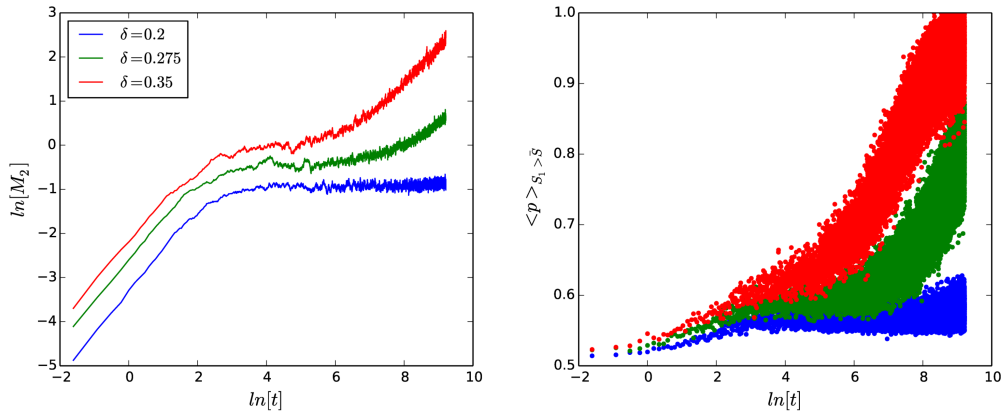

 \centering
 \includegraphics[width=\linewidth]
 {{{phenomenology_alpha0.3vary_delta_N1000_edited}}}
 \caption{(a) $M_2(t)$ on a log-log scale. (b) Probability $p$ (averaged over preceding 100 fights, considering only those fights in which $S_1>\bar{S}$) as a function of time. $\alpha=0.3$ for all curves.  $N=1000$, $n_r=10$.}
 \label{fig:phenom_M2_fixed_alpha}
\end{figure}

\subsection{Proof that a single individual with non-negligible status remains in the totalitarian end-state}
\label{section:SI_proof_one_indiv_endstate}

\paragraph{} When $\alpha > 0$, the system approaches an end-state in which a single individual possesses virtually all of the society's status and all other individuals have status approaching zero. An argument that only a single individual with non-negligible status remains in the end-state (and not, for example, multiple high-status individuals), is as follows.

\paragraph{} Consider a large (but finite) system of $N$ individuals. When $\alpha>0$, beginning from an egalitarian initial condition, the system evolves such that there are relatively few individuals with large status and relatively many individuals with small status (see section \ref{section:SI_phenom}). Since the average status of the system is constant, we can use it as a reference point to define ``large status" to mean greater than average status, and ``small status" to mean less than average status. The subsequent dynamics can be understood in terms of two timescales. The first timescale, $\lambda_1$, sets the time (measured in number of fights) that transpires, on average, between fights involving the same two particular individuals. Since pairs of individuals are picked randomly from among the population, $\lambda_1$ is determined by the system size $N$ and remains unchanged as the system evolves.

\paragraph{} The second timescale, $\lambda_2$, relates to the time required for an individual with less than average status to achieve an upset victory over an individual with greater than average status. Once the system has evolved to a point in which there are few individuals with high statuses and many individuals with low statuses, such an upset is the only way in which an individual with lower than average status can obtain greater than average status. $\lambda_2$ is determined by the average probability, $\langle p_{upset}\rangle$, with which an upset occurs (the smaller $\langle p_{upset}\rangle$, the larger $\lambda_2$). The probability for such an upset to occur (in any particular fight) is less than 50$\%$, by Eq 1 of the main text. Therefore, fights that involve one individual with greater than average status and one individual with less than average status will generally result in a transfer of status from the lower-status individual to the higher-status individual. This creates a net transfer of status from individuals with lower than average status to individuals with greater than average status. One outcome of this net transfer of status is a reduction in $\langle p_{upset}\rangle$. Consequently, the timescale $\lambda_2$ must increase with time (number of fights). 

\paragraph{} Since $\lambda_1$ remains constant whereas $\lambda_2$ increases as the system evolves, eventually $\lambda_1$ dominates the dynamics such that any two particular individuals fight one another much more frequently than any one particular individual with lower than average status achieves an upset that transforms it into an individual with greater than average status. Since fights between any two particular high-status individuals occur with higher frequency than fights in which a low-status individual achieves an upset over a high-status individual, there is a net flow of the losers of fights between two high-status individuals into the ranks of those with lower than average status. 

\paragraph{} The rate at which this net flow of losers occurs increases as $\lambda_2$ increases, and after an infinite number of fights ($t \to \infty$), only a single individual with greater than average status remains. 

\subsection{Technical aspects regarding $\tau_2$ extraction and errors}
\label{section:SI_errors_tau2}

\paragraph{} For small values of $\alpha$, $\tau_2$ diverges, such that very long simulations are required in order to evaluate $\tau_2$. In order to simplify matters, we consider a linear expansion of Eq  6 in the region where $\tau_1 \ll t \ll \tau_2$: 

\begin{equation}
M_2 = c_1 + (c_2-c_1)(1-e^{-t/\tau_2})
\approx c_1 + \frac{c_2-c_1}{\tau_2}t.
\end{equation} 

A linear fit to $M_2(t)$ data in the $\tau_1 \ll t \ll \tau_2$ can then be made, with intercept $c_1$ and slope $s = (c_2-c_1)/\tau_2$. This allows $\tau_2$ to be calculated, since $c_2$ is known from Eq 5.

\begin{figure}[H]
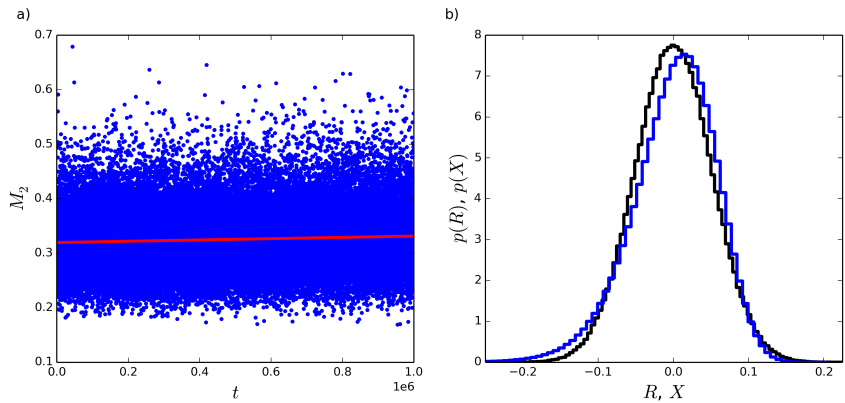

 \centering
 \includegraphics[width=\linewidth]
 {{{Residuals_and_M2_alpha0.2_delta0.2}}}
 \caption{(a) 10 overlaid realizations of $M_2(t)$ for $\delta=0.2$, $\alpha=0.2$, $N=100$, $n_r=10$, and linear fit (red line). (b) Distribution of residuals $R=y_{fit}-M_2$ (blue curve) and a Normal distribution with width equal to $\langle \sigma(M_2(t)) \rangle$ of a random variable $X$ (black curve).}
 \label{fig:SI_Residual_alpha0.2}
\end{figure}

\paragraph{} An example is shown in Fig  \ref{fig:SI_Residual_alpha0.2}a. Here, a set of 10 realizations of the simulation were performed. The plot contains the 10 different $M_2(t)$ curves overlayed one top of each other, with a linear fit (red line) to the combined data for all 10 curves. Fig  \ref{fig:SI_Residual_alpha0.2}b shows the distribution of residuals. The residuals are skewed, such that large increases in $M_2$ are more likely than large decreases. 

\paragraph{} Propagation of errors shows that $\Delta \ln[N/\tau_2] \approx \Delta s/s$, assuming that $\Delta s \gg \Delta c_1$ (justified below). This means that as $\alpha$ is decreased and $s$ approaches 0, it eventually becomes impossible to obtain a meaningful fitted value of $\tau_2$. Following this reasoning, we include points on the plot of $\ln[N/\tau_2]$ vs. $\alpha$ shown in Fig 7 for values of $\alpha$ for which $\Delta s/s < 1$.  

\paragraph{} Since $M_2(t)$ has a skewed (non-Normal) distribution of residuals and contains time-correlations, it is difficult to obtain a meaningful value of $\Delta s$ from a regression. We therefore estimate $\Delta s$ using uncorrelated, normally-distributed synthetic data with a width that is representative of the fluctuations in $M_2(t)$. This is done by generating random data from a Normal distribution with width equal to the average (over all time points) of the standard deviation taken, at each time $t$, across the 10 realizations of the simulation. An example of this synthetic data is shown in Fig \ref{fig:SI_Residual_alpha0.2}b (black curve). A least squares fit provides a correct determination of the error on the slope and intercept of this synthetic data, which we use as estimates for $\Delta s$ and $\Delta c_1$. In this way, we see that $\Delta s \gg \Delta c_1$.

\paragraph{} This procedure provides an estimate of the (small $\alpha$) limit beyond which points cannot be added to the plot of Fig 7 (obtaining further points would require longer simulation times that are beyond the scope of this study). The procedure does not provide a reasonable estimate of error bars for Fig 7. However, $\Delta\ln[N/\tau_2]$ is expected to increase with decreasing $\alpha$ (since $s$ decreases, while $\Delta s$ remains approximately constant), and this is reflected in increased fluctuations of $\ln[N/\tau_2]$ about the straight black lines in Fig 7 as $\alpha$ decreases.

\subsection{Equi-$\sigma$ distributions}
\label{section:SI_equi_sigma_distns}

\begin{figure}[H]
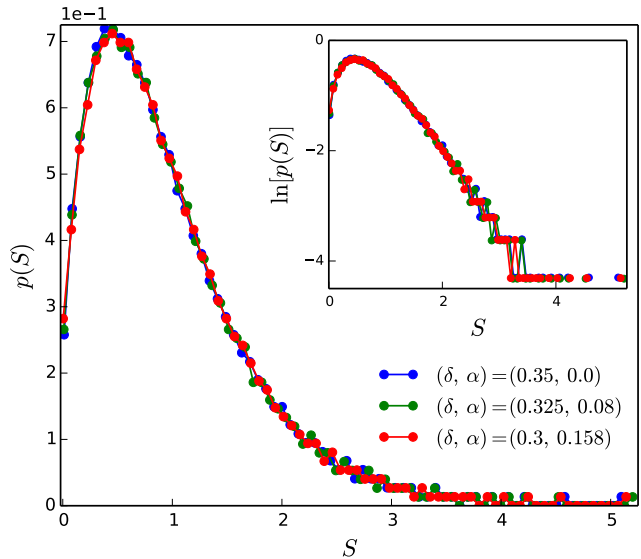

 \centering
 \includegraphics[scale=0.65]
 {{{equi_sigma_distns_N1000_nr500}}}
 \caption{Equi-$\sigma$ status distributions, obtained at $\tau_{obs}=1000$, for different pairs of ($\delta,\alpha$) values. Blue curve: $\left\langle \sigma \right\rangle = 0.733 \pm 0.001$; green curve: $\left\langle \sigma \right\rangle = 0.732 \pm 0.001$; red curve: $\left\langle \sigma \right\rangle = 0.732 \pm 0.001$. $\left\langle \mu_3/\sigma^3 \right\rangle = 1.46 \pm 0.01$ for all three distributions.}
 \label{fig:equi_sigma_distns}
\end{figure}

\paragraph{} Essentially identical distributions can be produced (within a particular simulation time) for different combinations of the parameters $\delta$ and $\alpha$. This is shown in Fig \ref{fig:equi_sigma_distns}, which plots status distributions generated for three different points in $\delta-\alpha$ parameter-space. The inset to Fig \ref{fig:equi_sigma_distns} shows the distributions with a logarithmic scale on the y-axis, in order to allow closer inspection of the tails of the distributions. To compare the shapes of the distributions in Fig \ref{fig:equi_sigma_distns}, in addition to visual inspection of the plotted curves, we use the time-and-realization-averaged standard deviation, $\left\langle \sigma \right\rangle$ of the status distributions. This is defined as follows: first, we average the standard deviation of the status distributions over 500 realizations of the simulation, giving us $\sigma(t)$. Then, we average $\sigma(t)$ over a range of time ($500 \leq t \leq 1000$) during which $\sigma(t)$ is approximately constant corresponding to the long-lived state. This gives us $\left\langle \sigma \right\rangle$. Repeating this procedure for the skewness of the status distributions gives us $\left\langle \mu_3/\sigma^3 \right\rangle$, which we also use to compare the shapes of the distributions in Fig \ref{fig:equi_sigma_distns}.
 
\paragraph{} The value of $\left\langle \sigma \right\rangle$ for the three distributions is the same, within error (caption of Fig \ref{fig:equi_sigma_distns}). The error value used here is the standard deviation of $\sigma(t)$ taken over the range of time $500 \leq t \leq 1000$. Likewise,  $\left\langle \mu_3/\sigma^3 \right\rangle$ is the same, within error, for the three different distributions. There is, however, one subtle difference between the distributions that can be seen in Fig \ref{fig:equi_sigma_distns}: the value of $p(S)$ at $S=0$ becomes systematically larger going from smaller (larger) to larger (smaller) $\alpha$  ($\delta$). Except for this caveat, Fig \ref{fig:equi_sigma_distns} shows the existence of sets of ($\delta, \alpha$) points producing essentially identical distributions.

\subsection{Determining the location of the transition between regions II and III of Fig 10}
\label{section:SI_transition_location}

\subsubsection{Using a criterion based on the slope of $M_2(t)$}
\label{section:SI_transition_slope_criterion}

\paragraph{} The location of this transition between regions II and III of Fig 10 depends on the time, $\tau_{obs}$, over which the system is observed (simulated). A simple criterion equates the onset of runaway with the appearance of a positive slope in the long-time portion of $M_2(t)$. Whether such a slope is observed in $M_2(t)$ depends on the size of the fluctuations in the simulation data. We ran $n_r=25$ realizations of the simulation for system size $N=1000$. In order to determine if the system is in the long-lived state for a particular value of $\alpha$, we first fit a straight line to $M_2(t)$ over a range [$t_0$, $\tau_{obs}$], where $t_0 \gg \tau_1$ such that $M(t_0)$ is on the $M_2(t)$ plateau that follows the initial transient evolution away from the initial (egalitarian) distribution. The transition is considered to have occurred at the value of $\alpha$ for which the long-time part of the $M_2(t)$ curve increases by $2.5\%$ over the observation (simulation) time $\tau_{obs}$. For the choice of $N$ and $n_r$ used, this amount of increase corresponds to approximately 5 times the standard deviation of the residual $R(t) = \hat{M_2}(t) - M_2(t)$, where $\hat{M_2}(t)$ is the linear fit, and $M_2(t)$ is averaged over the $n_r=25$ realizations. Therefore, the distribution is considered to be long-lived over the interval [$t_0$, $\tau_{obs}$] as long as $m(\tau_{obs}-t_0)<0.025b$, where $m$ is the slope and $b$ the intercept of the linear fit.

\subsubsection{Using the Arrhenius relation between $\tau_2$ and $\alpha$} 
\label{section:SI_transition_Arrhenius_relation}

\paragraph{} The criterion (see above, section \ref{section:SI_transition_slope_criterion}) of a $2.5\%$ increase in $M_2(t)$, used to determine the location of the transition between the long-lived state and runaway, can be written in terms of Eq 6 as:

\begin{equation} 
\label{Eq:M2_oper_criterion}
M_2(\tau_{obs}) = 1.025c_1 \approx c_1 + (c_2-c_1)(1-e^{-\tau_{obs}/\tau_2}), 
\end{equation}

where the approximation comes from the assumption that $\tau_{obs} \gg \tau_1$. Rearranging Eq. \ref{Eq:M2_oper_criterion} in terms of $\tau_{obs}$ gives the following relationship: 

\begin{equation} 
\label{Eq:tau_obs}
\tau_{obs} \approx -\tau_2\ln\left(1-\frac{0.025c_1}{N-1-c_1}\right).
\end{equation}

\paragraph{} Now, we fix $\tau_{obs}$ and adjust $\alpha$ until the relationship in Eq. \ref{Eq:tau_obs} is satisfied. Inserting the relationships $\alpha_b = 0.53\delta^{-1.21}$ (Fig 8) and $Nf_0 = \delta^{1.3}$ (Fig 9) into Eq 7 and rearranging, we find: 

\begin{eqnarray}
\label{Eq:alpha_t}
\alpha_t & \approx & \frac{\alpha_b}{\ln[\tau_2(\alpha_t)f_0]} \nonumber \\
\nonumber \\
& \approx & \frac{0.53\delta^{-1.21}}{\ln\left[\frac{-\tau_{obs}\delta^{1.3}}{N\ln\left(1-\frac{0.025c_1}{N-1-c_1}\right)}\right]},
\end{eqnarray}

where $\alpha_t$ is the location of the transition  (as defined by the criterion stated above) and  $\tau_2(\alpha_t)$ is the value of $\tau_2$ when $\alpha = \alpha_t$.

\paragraph{} A factor of $c_1$ appears twice in Eq. \ref{Eq:alpha_t}. While $c_1$ is in fact a function of $\delta$ and $\alpha$, it changes slowly with $\alpha$ (Fig 6d). Therefore, we set $c_1$ to be equal to its value when $\alpha=0$, such that $c_1=\delta/(1-\delta)$ (see Eq 2). With these assumptions in place, Eq. \ref{Eq:alpha_t} produces the solid black ($\tau_{obs}=10^4$) and red ($\tau_{obs}=10^3$) curves in Fig 10.

\subsubsection{$N$-dependence of $\alpha_t$}

\begin{figure}[H]
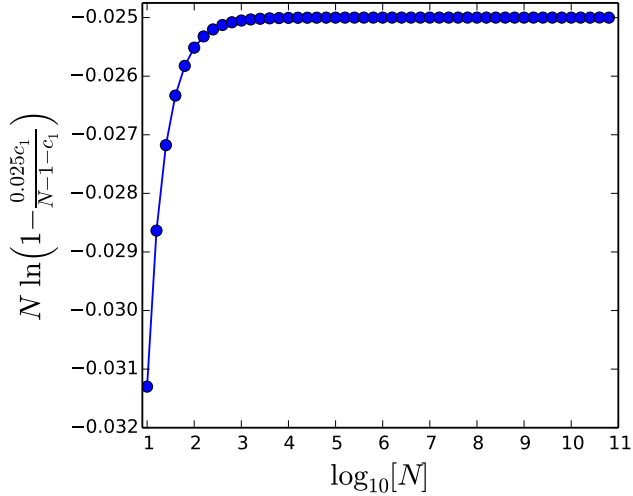

 \centering
 \includegraphics[scale=0.7]
 {{{N_dependence_of_alpha_t}}}
 \caption{The part of Eq. \ref{Eq:alpha_t} containing a dependence on $N$ plotted vs. powers of $N$, and becomes effectively constant at large $N$.} 
 \label{fig:SI_Ndependence_alphat}
\end{figure}

\paragraph{} The $N$-dependence of $\alpha_t$ is contained within the denominator of the argument of the logarithm in the denominator of Eq. \ref{Eq:alpha_t}. Fig \ref{fig:SI_Ndependence_alphat} shows a plot of this part of Eq. \ref{Eq:alpha_t} as a function of $N$. The plot becomes effectively constant for large $N$, showing that the $N$-dependence of $\alpha_t$ vanishes for increasing $N$ such that the location of the transition between the long-lived state and runaway effectively does not depend on system size.

\subsubsection{Alternative location of the transition between regions II and III according to Arrhenius relation}

\paragraph{} In some physical systems governed by an Arrhenius relation, the location of the operational transition between two regimes is taken by the researcher to be at the value of the control parameter (usually temperature, in physical systems) for which the transition rate $\tau$ is equal to the measurement time $\tau_m$. This is the case in the blocking transition of superparamagnetism, for example \cite{Knobel2008}. 

\paragraph{} We could use this reasoning to locate the transition between the long-lived state and runaway in our model at a value of $\alpha = \alpha_t^*$ for which $\tau_2 = \tau_{obs}$. Re-arranging Eq. 7 and substituting in the relationships for $\alpha_b$ and $Nf_0$, we find: 

\begin{eqnarray}
\label{Eq:alpha_t*}
\alpha_t^* & = & \frac{\alpha_b}{\ln[\tau_{obs}f_0]} \nonumber \\
\nonumber \\
& = & \frac{0.53\delta^{-1.21}}{\ln\left[\frac{1.03\delta^{1.3}\tau_{obs}}{N}\right]}.
\end{eqnarray}

\paragraph{} In a physical experiment, the criterion for setting the location of the operational transition is primarily determined by the measurement type. For some measurement types, it may be appropriate or natural to set the location of the transition at a value of the control parameter analogous to $\alpha_t^*$. In our system, this would correspond to identifying the transition between the long-lived state and runaway at a value of $\alpha$ for which $M_2 = M_2(t=\tau_2) \approx c_2(1-1/e)$. This represents a significant evolution of the system toward the totalitarian end-state, which is not suitable for our purpose. Rather, our goal is to determine the value of $\alpha$ for which, on the scale of the observation time considered, the long-lived state plateau value of $M_2$ is lost and a noticeable increase in $M_2(t)$ is observed. We therefore place the location of the transition at $\alpha_t$.

\newpage
\subsection{Additional details about evolution of extended model distributions}
\label{section:SI_extended_model_evolution}

\paragraph{} The figures in this section are included to show how the status distributions produced by the extended model (presented in section 2.1 of the main text) evolve in time. In the upper left panel of each figure, the evolution of $M_2(t)$ on a log-log scale is shown. The status distributions are shown at various points in time (indicated within the plots) in the remaining panels. For each figure, the simulation had system size $N=1000$, and the results are averaged over $n_r=5$ realizations of the simulation. The simulation parameters $\delta=0.2$, $\eta=1.0$, and $\epsilon=0.1$ are fixed for all figures, and the parameter $\alpha$ is varied.

\begin{figure}[H]
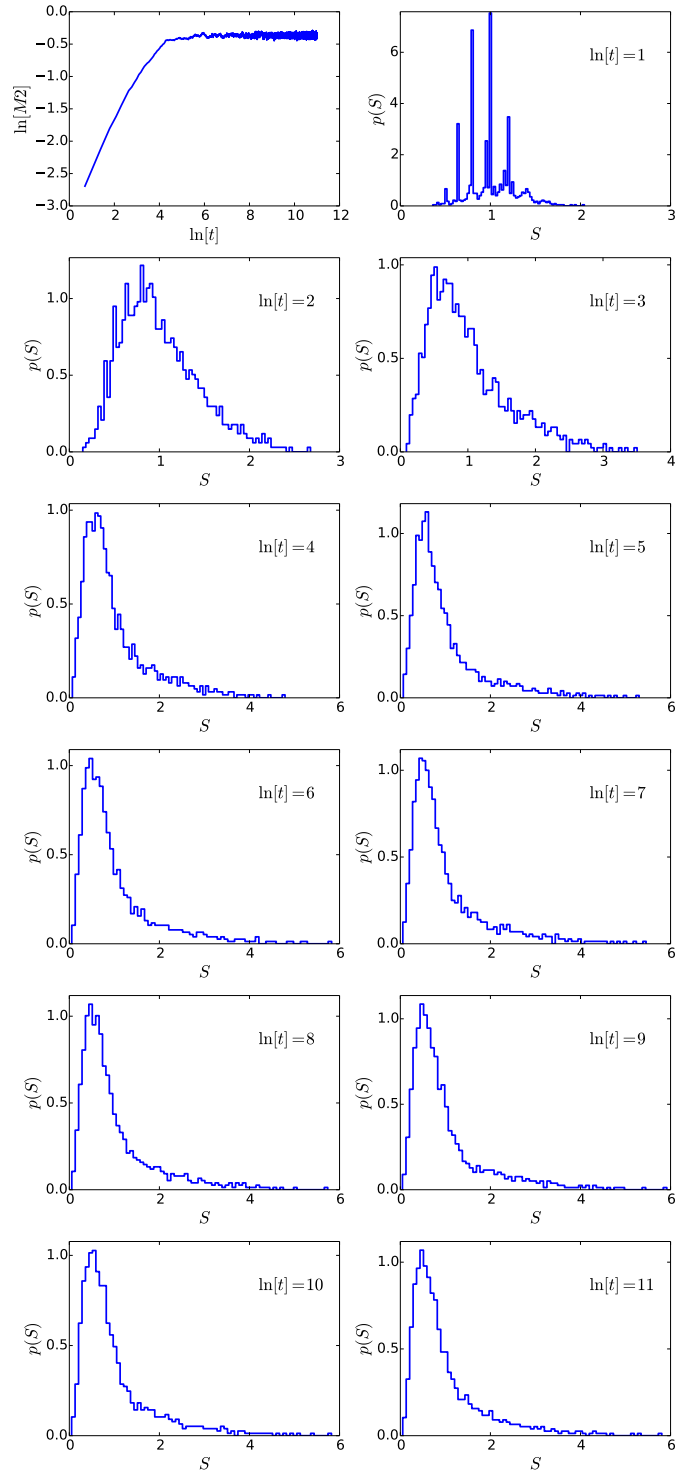

 \centering
 \includegraphics[height=\textheight]
 {{{SI_diff_model_evoln_hists_delta0.2_alpha0.0_eta1.0_eps0.1}}}
 \caption{Time evolution of distributions for $\delta=0.2$, $\alpha=0.0$, $\eta=1.0$, $\epsilon=0.1$.}
 \label{fig:hists_v_time_alpha0.0_eta1.0_eps0.1}
\end{figure}

\begin{figure}[H]
 \centering
 \includegraphics[height=\textheight]
 {{{SI_diff_model_evoln_hists_delta0.2_alpha0.2_eta1.0_eps0.1}}}
 \caption{Time evolution of distributions for $\delta=0.2$, $\alpha=0.2$, $\eta=1.0$, $\epsilon=0.1$.}
 \label{fig:hists_v_time_alpha0.2_eta1.0_eps0.1}
\end{figure}

\begin{figure}[H]
 \centering
 \includegraphics[height=\textheight]
 {{{SI_diff_model_evoln_hists_delta0.2_alpha0.4_eta1.0_eps0.1}}}
 \caption{Time evolution of distributions for $\delta=0.2$, $\alpha=0.4$, $\eta=1.0$, $\epsilon=0.1$.}
 \label{fig:hists_v_time_alpha0.4_eta1.0_eps0.1}
\end{figure}

\begin{figure}[H]
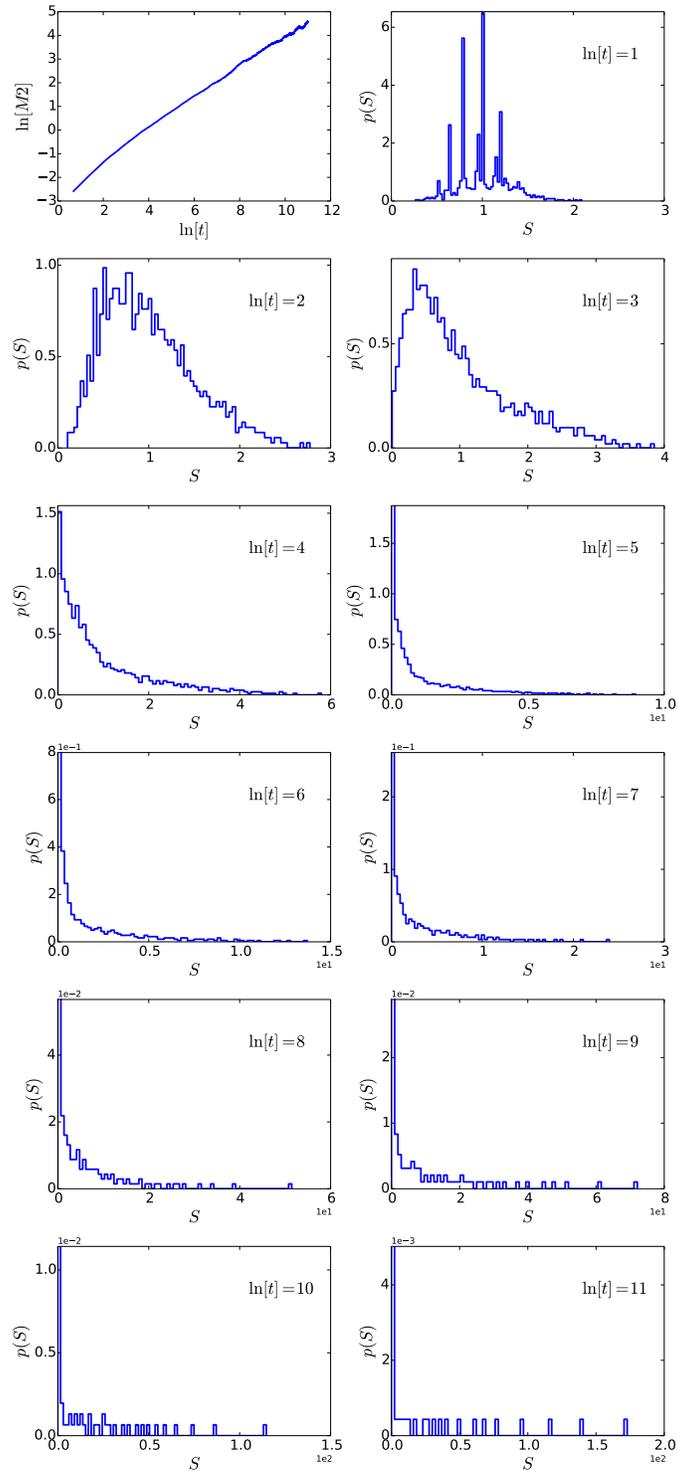

 \centering
 \includegraphics[height=\textheight]
 {{{SI_diff_model_evoln_hists_delta0.2_alpha0.75_eta1.0_eps0.1}}}
 \caption{Time evolution of distributions for $\delta=0.2$, $\alpha=0.75$, $\eta=1.0$, $\epsilon=0.1$.}
 \label{fig:hists_v_time_alpha0.75_eta1.0_eps0.1}
\end{figure}

\newpage
\setcounter{subsection}{0}
\setcounter{figure}{0}
\renewcommand{\thesubsection}{S2.\Alph{subsection}}
\renewcommand{\figurename}{Fig}
\renewcommand\thefigure{S2.\arabic{figure}}
\renewcommand\thetable{S2.\arabic{table}}

\section*{Appendix S2} \label{S2_Appendix}
\addcontentsline{toc}{section}{Appendix S2}

\subsection{Dependence of interaction probabilities on rank for mountain goats in study of C{\^{o}}t{\'{e}} (2000)}

\paragraph{} In section 4.1 of the main text, we make use of data from the study of C{\^{o}}t{\'{e}} \cite{Cote2000} on agonistic interactions in female mountain goats collected during four consecutive summers. To compare results from our simulations with the mountain goat data across multiple years, we consider only those individual's in C{\^{o}}t{\'{e}}'s datasets that are present in all four years. This is a subset of the full dataset, since some mountain goats enter the society (e.g. by ageing to the age of three years old) and leave the society (e.g. by dying) from one year to the next.

\paragraph{} To examine whether the mountain goats in our subset taken from C{\^{o}}t{\'{e}}'s data had a tendency to interact more frequently with those close in rank versus those far away in rank, we used the same approach as was used in Fig 6 of Ref. \cite{Cote2000}. That is, for each individual, we considered the percentage of the individual's interactions with the 10 individuals closest in rank and with the 10 individuals furthest in rank. For the $N=26$ individuals present in all 4 years of the study, there was no bias toward interacting with individuals close in rank or far in rank. Results are presented in the following table. 

\begin{table}[h!]
 \caption{Tendency of individuals to interact with other individuals close or far in rank in $N=26$ subset from C{\^{o}}t{\'{e}}'s mountain goat data. Each entry in columns 2-4 shows the percentage (averaged over the $N=26$ individuals) of an individual's interactions that were against (column 2) one of the 10 closest ranked other individuals; (column 3) one of the 10 furthest ranked other individuals; or (column 4) an individual not belonging to the group of 10 closest to or 10 furthest from the individual in rank. Each error value shows the standard deviation of the percentage of interactions.}
 \label{tab:CoteInteractions}
 \begin{center}
 \begin{tabular}{l|c|c|c}
Year & 10 closest & 10 furthest & Other \\
\hline
1994 & 0.39 $\pm$ 0.09 & 0.4 $\pm$ 0.1 & 0.2 $\pm$ 0.1\\
1995 & 0.4 $\pm$ 0.2 & 0.4 $\pm$ 0.1 & 0.2 $\pm$ 0.1\\
1996 & 0.4 $\pm$ 0.1 & 0.4 $\pm$ 0.1 & 0.2 $\pm$ 0.1\\
1997 & 0.4 $\pm$ 0.1 & 0.4 $\pm$ 0.2 & 0.19 $\pm$ 0.09\\

 \end{tabular}
 \end{center}
\end{table}

\paragraph{} Similar results are obtained when considering the 5 closest and 5 furthest individuals in rank and the 15 closest and 15 furthest individuals in rank.

\newpage
\subsection{Measurable quantities that may serve as proxies for status in non-human animals}

\begin{center}
\begin{longtable}{| L{.20\textwidth} | L{.80\textwidth} |}
\caption{Measurable quantities that are relevant to the estimation of status proxies}\\
\hline
Agonistic behaviour (fights) & 
\begin{itemize}
\item Agonistic interactions (fights) between pairs of individuals are observed in many studies and used to assign a rank to individuals in a dominance hierarchy. These can include physical aggressions, physical and non-physical (e.g. vocal, eye-contact) intimidations, and subordinations \cite{Sapolsky2005, Hsu2006, Hardy2013, Shizuka2015}.
\item Typically, an interaction matrix is constructed, with one row and column for each observed individual, where the entries are the numbers of times that individual $i$ has defeated individual $j$ in a fight \cite{Albers2001, Neumann2011, SanchezTojar2018}.
\item A ranking of the individuals in the study can be obtained by re-arranging the rows of the interaction matrix to satisfy a criterion (e.g. minimizing the sum of entries below the diagonal) \cite{DeVries1998} or by calculating a score from the interaction matrix (e.g. David’s Score) \cite{David1987}.
\item Other studies make use of the sequence of interactions to rank individuals using a score that evolves each time an individual has an interaction. With these methods, the rank-ordering of the individuals can change as more interactions occur \cite{Albers2001, Neumann2011, SanchezTojar2018, NewtonFisher2017}.
\end{itemize} \\
\hline 
Affiliative behaviours (grooming in primates) &
\begin{itemize}
\item The grooming time that an individual receives correlates with hierarchical position (grooming is directed “up the hierarchy”) \cite{Tiddi2012, SnyderMackler2016}, and may be given as a service in exchange for support in agonistic interactions or tolerance \cite{Schino2006, Seyfarth1977}. However, individuals have limited grooming-time to give, and compete to groom higher-ranking individuals \cite{Schino2001}. Therefore, grooming-time received by a high-status individual may underestimate her status.
\end{itemize} \\ 
\hline
Physical characteristics & 
\begin{itemize}
\item Whereas animals living in small groups may be able to assess the fighting abilities of one another by remembering the history of past interactions, when these same animals live in large groups (e.g. of $N \approx 100$ \cite{DEath2003}), in which the role of individual recognition is reduced and where it is unlikely that every individual has interacted with every other individual, they may instead use physical characteristics (so-called "status signals") to communicate and assess fighting ability.
\item In some cases, the status signal can change on a short time-scale, such as following a promotion or demotion in rank, such that it may signal the current fighting ability of the individual to others (e.g. the intensity of red colour in geladas and mandrills).
\item Examples: Black facial marks in wasps \cite{Tibbetts2008, Tibbetts2009};  comb size in hens \cite{DEath2003}; dark plumage throat-patch size in male sparrows \cite{Moller1987}; red chest-patch colour in male geladas \cite{Bergman2009}; red face colour in male mandrills \cite{Setchell2008}.
\end{itemize}
\\ 
\hline
Biochemical concentrations (blood and saliva chemistry) & 
\begin{itemize}
\item Hormone and neurotransmitter concentrations are related to position in the dominance hierarchy and the dominance behaviour of the individual \cite{Eisenegger2011, McCall2012, Sapolsky2005}. However, the relationship between biochemical concentrations and hierarchical rank is complicated in that it depends on social context. For example: 
\item High ranking individuals may have elevated levels of glucocorticoid (stress hormone), but for different reasons than low-ranking individuals with the same elevated level of glucocorticoid \cite{Gesquiere2011, Sapolsky2011}. 
\item Testosterone concentration correlates with rank in the dominance hierarchy (higher-ranking individuals have higher concentrations of testosterone), but only in periods of societal instability \cite{Setchell2008, Eisenegger2011}.
\item Serotonin concentration correlates with rank, but is very sensitive to the presence of subordinates \cite{Watanabe2015}.
\end{itemize} \\ 
\hline
Body size (such as weight or length) & 
\begin{itemize}
\item Correlated with dominance rank and ability to win agonistic encounters in social insects, including ants, bees, and wasps \cite{Hardy2013, Tibbetts2008}, crustaceans \cite{Adams1990}, fish \cite{Beaugrand1996, Buston2003}, and reptiles \cite{Schuett1997}, although there are confounding factors such as age \cite{Hughes1988} and past fighting experience \cite{Withee2016, Rutte2006}.
\item Less strongly related to dominance rank in more "complex" animals such as mammals \cite{Pusey1997, Paoli2006, Cote2000}.
\end{itemize} \\ 
\hline
Age & 
\begin{itemize}
\item Dominance rank is correlated with age in many species \cite{Hughes1988, Higashi1994, Thouless1986}, although confounding factors include past fighting experience and body size \cite{Hughes1988}.
\item In many primates, the alpha eventually loses his/her position to a younger challenger \cite{Setchell2010, Zhao2009, Uehara1994, Sapolsky1993}, such that a simple linear relationship between age and dominance hierarchy rank is not accurate.
\end{itemize} \\
\hline
Preferential access to food & 
\begin{itemize}
\item Access to high quality food is correlated with dominance rank in many species, for example: salmon \cite{Maclean2001}, caribou \cite{Barrette1986}, deer \cite{Appleby1980}, goats \cite{Barroso2000}, macaques \cite{Sterck1997}, chimpanzees \cite{Wittig2003}, and baboons \cite{Marmot2014}.
\item Dominance not correlated with access to food in carrion crows \cite{Richner1989}.
\end{itemize} \\
\hline
Mating opportunities and reproduction & 
\begin{itemize}
\item Access to reproductive opportunities generally correlates with position in the dominance hierarchy \cite{Cowlishaw1991, Ellis1995}, however the relationship can be non-linear. For example, high ranking female baboons have been found to have more miscarriages and a higher rate of infertility \cite{Packer1995}, beta-male chimpanzees in some groups have much fewer copulations than lower-ranking males \cite{NewtonFisher2004}, and low-ranking male monkeys with affiliative relationships with females can have greater reproductive success than expected given their rank \cite{Sapolsky2005}. 
\end{itemize} \\
\hline
\end{longtable}
\end{center}

\newpage

\subsection{Sensitivity of fit to USA income data to change in parameter $\epsilon$}
\label{section:SI_extended_sensitivity}

\paragraph{} Fig 15 of the main text shows a fit of the extended model to the USA household income distribution. The parameter $\eta$ is set such that $\eta\bar{S} = S_B$, where $S_B$ is the ``break point" in the data. The parameter $\epsilon$ in Fig 15 was chosen to be equal to 0.08 in order to obtain a good fit to the income data. Two figures are included below to show how decreasing (Fig \ref{fig:USA_fit_eta3.5_eps0.06}) or increasing (Fig \ref{fig:USA_fit_eta3.5_eps0.1}) the value of $\epsilon$ affects the fit to the proxy data.

\begin{figure}[H]
 \centering
 \includegraphics[scale=0.75]
 {{{USA_fit_difference_model__delta0.4_alpha0.0_eta3.5_eps0.06}}}
 \caption{Fit of extended model ($\eta=3.5$, $\epsilon=0.06$) to 2015 USA household income data.}
 \label{fig:USA_fit_eta3.5_eps0.06}
\end{figure}

\begin{figure}[H]
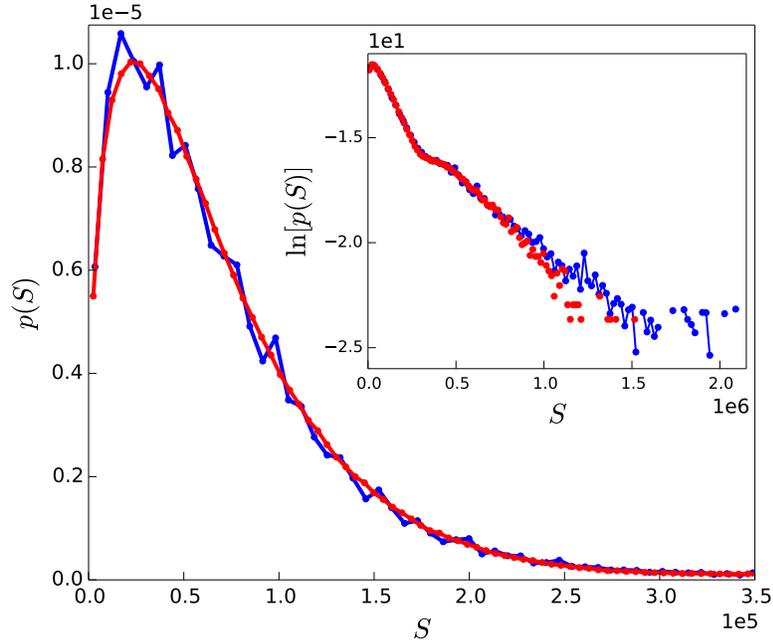

 \centering
 \includegraphics[scale=0.75]
 {{{USA_fit_difference_model__delta0.4_alpha0.0_eta3.5_eps0.1}}}
 \caption{Fit of extended model ($\eta=3.5$, $\epsilon=0.1$) to 2015 USA household income data.}
 \label{fig:USA_fit_eta3.5_eps0.1}
\end{figure}

\subsection{High-income tails of USA household income distributions}
\label{section:SI_proxy_distn_tails}

\paragraph{} In Fig 16 of the main text, fits of exponential and power-law distributions to the high-income tail of the 2000 and 2015 USA household income distributions were shown. Normalized with respect to the boundaries $S_l$ and $S_h$, the probability density function, $p(S)$, of the power-law distribution is: 

\begin{equation}
\label{Eq:PL_Sl_Sh}
p(S) = \frac{S_l^{1-\gamma}-S_h^{1-\gamma}}{\gamma-1} S^{-\gamma},
\end{equation}

and that of the exponential distribution is:

\begin{equation}
\label{Eq:EXP_Sl_Sh}
p(S) = \frac{e^{(S_l-S)/T}}{T(1-e^{(S_l-S_h)/T})},
\end{equation}

where the parameters $\gamma$ and $T$ are determined by maximum likelihood estimation using the data within the range $[S_l,S_h]$ \cite{Baro2012}. Two additional figures showing different $[S_l,S_h]$ ranges are included below. As for Fig 16 of the main text, the black dashed line shows the power-law fit and the solid red line shows the exponential fit. The 2015 curves have been shifted down in the plots for better visualization.

\begin{figure}[H]
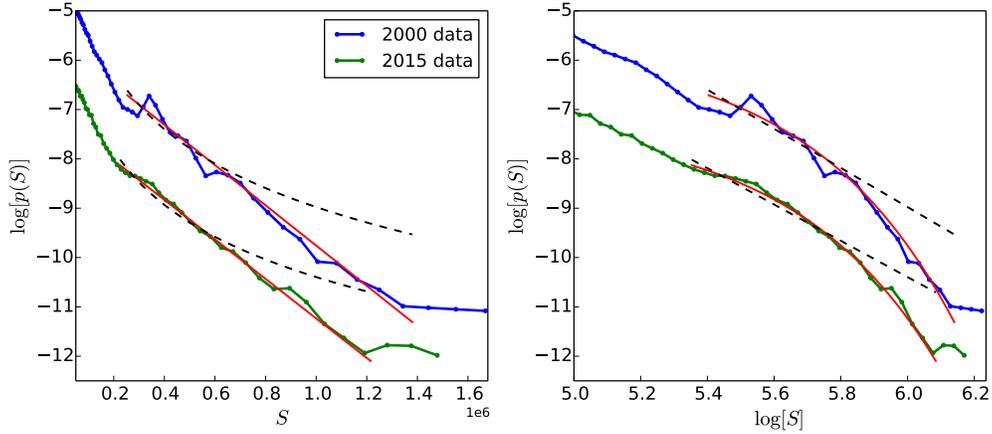

 \makebox[\textwidth][c]{\includegraphics[width=\textwidth]{{{PDF_EXP_PL_twopanel_partoftail}}}}
 \caption{Power-law (dashed black line) and exponential (solid red line) distributions with $S_l$ and $S_h$ chosen to correspond to the high-income tail, excluding the highest-income points. For the fits to the 2015 data shown in this figure, the choice of $S_h$ resulted in the exclusion of the 15 largest data points. For the fits to the 2000 data shown in this figure, the choice of $S_h$ resulted in the exclusion of the two largest data points. Exclusion of the highest-income data points is justified because high-income cutoffs have been artificially applied to the USA data by the governmental agency that provided it, for the purpose of protecting confidentiality. $S$ represents USA household income data in 1999 USD.}
 \label{fig:PL_EXP_fit_real_data_partoftail}
\end{figure}

\begin{figure}[H]
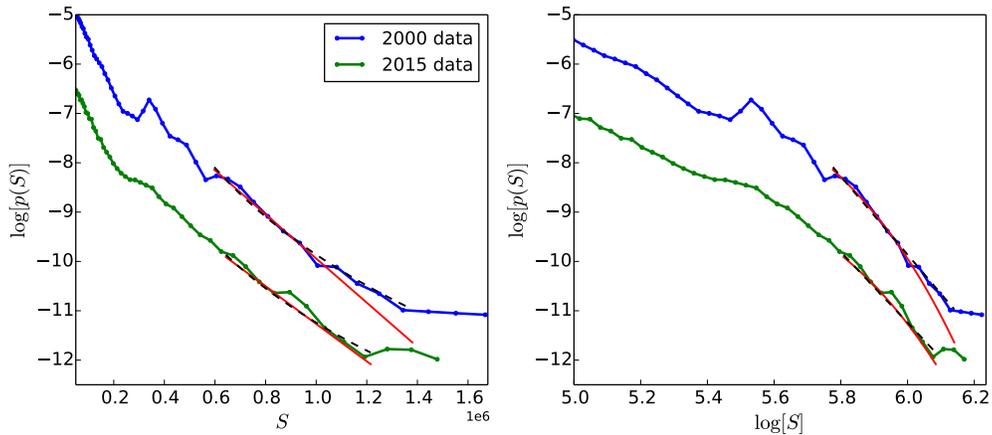

 \makebox[\textwidth][c]{\includegraphics[width=\textwidth]
 {{{PDF_EXP_PL_twopanel_endoftailcutoff}}}}
 \caption{Power-law (dashed black line) and exponential (solid red line) distributions with $S_l$ and $S_h$ chosen to correspond approximately to the latter part of the high-income tail. Graphical analysis shows that, at best, a power-law can only fit segments of the high-income tail. $S$ represents USA household income data in 1999 USD.}
 \label{fig:PL_EXP_fit_real_data_endoftailcutoff}
\end{figure}

\subsubsection{Kolmogorov-Smirnov (KS) test} 
\label{section:SI_KS_test}

\paragraph{} A Kolmogorov-Smirnov (KS) test was used to evaluate whether the high-income tail of the USA household income distribution is consistent with a power-law distribution (Eq \ref{Eq:PL_Sl_Sh}) or with an exponential distribution (Eq \ref{Eq:EXP_Sl_Sh}). The term ``theoretical distribution" is used below to refer to the distribution (either power-law or exponential) that the real data is compared to in the statistical test.

\paragraph{} The KS test relies on the ``KS distance", which is the largest distance between the cumulative distribution function (CDF) of the theoretical distribution (with specified parameters) and the empirical cumulative distribution function, $F_N$, determined directly from the data as follows: 

\begin{equation}
F_N(x) = \frac{1}{N}\sum_{i=1}^N I_{[-\infty,x]}(X_i),
\end{equation}

where $I_{[-\infty,x]}(X_i)$ is the indicator function, which is equal to $1$ if $X_i \le x$ and $0$ otherwise, and $N$ is the number of data points in the dataset. The 2015 USA household income dataset is a weighted dataset, such that every data point represents a particular number of people in the overall population. For this dataset, the weighted empirical cumulative distribution function \cite{SAS2000} was used: 

\begin{equation}
F_w(x) = \frac{1}{\sum_i w_i}\sum_{i=1}^N w_iI_{[-\infty,x]}(X_i),
\end{equation}

where $w_i$ is the weight assigned to the $i_{th}$ data point. 

\paragraph{} Since the parameters $\gamma$ (power-law distribution) and $T$ (exponential distribution) are unknown, they are estimated from the data using maximum likelihood estimation for the unweighted 2000 dataset and weighted maximum likelihood estimation for the 2015 dataset \cite{Wang2001}. 

\paragraph{} The KS distance, $D_{data}$, between the empirical CDF ($F_N$ or $F_w$) and the CDF of the theoretical distribution is compared to a distribution of KS distances determined from synthetic datasets. To obtain the latter distribution of KS distances, the following steps are repeated many times: 1) a sample of synthetic data containing the same number of data points as the real data is drawn from the theoretical distribution; 2) the parameters of the theoretical distribution are re-estimated from the synthetic data, in order to avoid biases that arise if this re-estimation is not performed \cite{Capasso2009}; 3) the KS distance, $D_{synth}$, is determined from the empirical CDF of the synthetic data and the theoretical distribution with re-estimated parameters. 

\paragraph{} If the real data is consistent with the theoretical distribution, $D_{data}$ should be smaller than a significant fraction of the set of $\{D_{synth}\}$. The fraction of $\{D_{synth}\}$ that is larger than $D_{data}$ is the $p$-value given by the test. We consider a $p$-value greater than 0.1 to be evidence of compatibility with the theoretical distribution. In other words, with a $p$-value greater than 0.1, there is not enough evidence to reject the null hypothesis that the real data comes from the theoretical distribution.

\paragraph{} A KS test showed that the high-income tail of the 2015 USA household income data is only compatible ($p$-value $> 0.1$) with an exponential distribution within particular ranges $[S_l,S_h]$ within the high-income tail, the largest of which is $\$700,000$ to $\$1,700,000$ ($\$492,800$ to $\$1,196,800$ in 1999 USD), and that the high-income tail is not compatible with a power-law for any of the ranges tested. The KS test for the 2000 USA household data also shows that the data is only compatible with an exponential distribution within particular ranges, the largest being from $\$800,000$ to the largest income value in the dataset ($\$1,668,400$), and that the high-income tail is not compatible with a power-law for any of the ranges tested.

\subsection{Alternative ``extended" models that pre-suppose a two-class structure}
\label{section:SI_alternative_models_two_class}

\subsubsection{Addition to extended model to allow fights between high-status individuals}
\label{section:SI_3condition_model}

\begin{figure}[H]
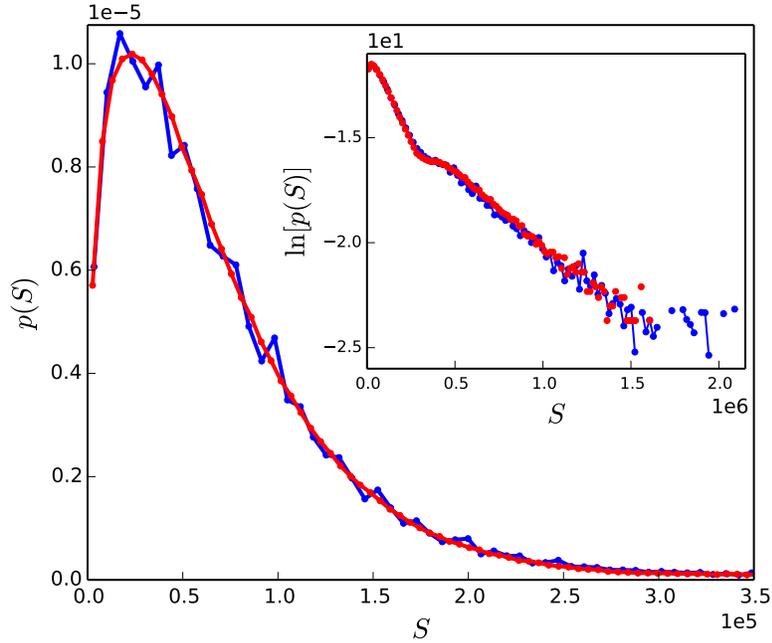

 \centering
 \includegraphics[scale=0.75]
 {{{USA_fit_3condition_model__delta0.4_alpha0.0_eta3.5_eps0.08}}}
 \caption{Fit of 3-condition extended model status distribution (red curve) to the 2015 USA household income distribution (blue curve). Simulation parameters: $\delta=0.4$, $\alpha=0$, $\eta=3.5$, $\epsilon=0.08$.}
 \label{fig:USA_fit_3condition}
\end{figure}

\paragraph{} In the extended model presented in section 2.1 of the main text, individuals at the top-end of the status distribution are separated by large amounts of status, typically greater than the amount $\eta\bar{S}$. These high-status individuals are therefore prevented from fighting with each other by the first condition ($S1-S2 > \eta\bar{S}$) of the extended model. This excludes many fights between high-status ``dominant" individuals and higher-than-average status ``challengers", whereas such dominant-challenger fights are important and common in real dominance hierarchies \cite{Gesquiere2011, Sapolsky2011, Sapolsky1992}. In order to allow high-status individuals to fight each other more frequently, we introduce a third condition to the extended model, such that the fight occurs if the statuses of both potential competitors are greater than the threshold amount $\eta\bar{S}$ (equivalently, if $S_2 > \eta\bar{S}$, since $S_1 >= S_2$). Under this modification to the extended model, the fight occurs if $S_1-S_2 \leq \eta\bar{S}$ OR $S2 > \eta\bar{S}$ OR $r \leq \epsilon$, where $r$ is a random number between 0 and 1. The additional condition $S2 > \eta\bar{S}$, pre-supposes a two-class structure \textit{a priori}. The status distributions produced by this 3-condition extended model have large-$S$ tails that decay exponentially over the full extent of the tail, unlike those produced by the 2-condition extended model presented in section 2.1 of the main text, which show a cutoff at very high values of $S$ (see Fig 12f of the main text). The status distributions of the 3-condition model therefore show improved fits to the proxy data, as can be seen in Fig \ref{fig:USA_fit_3condition}.

\subsubsection{Combining two independently simulated status distributions}
\label{section:SI_two_populations}

\begin{figure}[H]
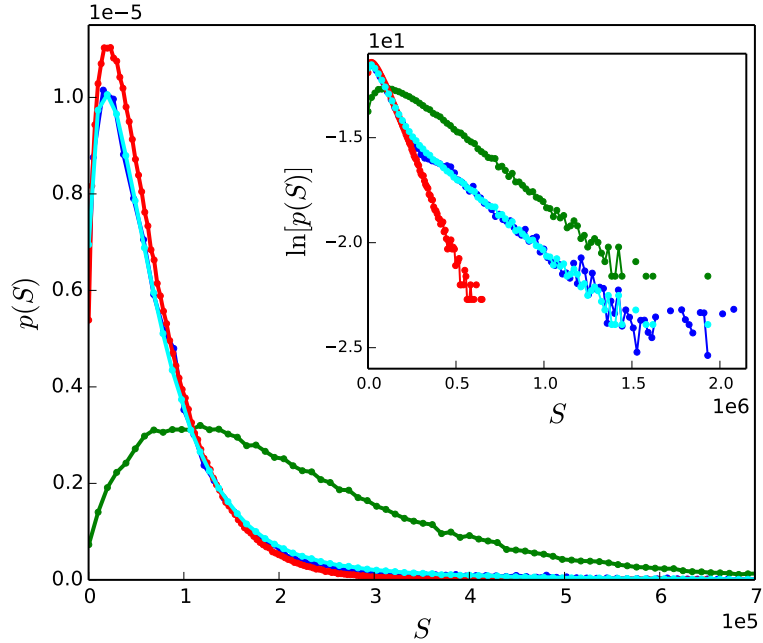

 \centering
 \includegraphics[scale=0.75]
 {{{USA_fit_2populations_orig_model}}}
 \caption{Fit of simulated distribution to USA data. Cyan curve is the combination of red and green distributions. For red curve, $N=0.9N_{dat}$, $\bar{S}=64,445$, and $\delta=0.4$; for green curve, $N=0.1N_{dat}$, $\bar{S}=225,000$, and $\delta=0.35$. $\alpha=0$ for both red and green curves.}
 \label{fig:USA_fit_2populations}
\end{figure}

\paragraph{} Here, we have used the original (two-parameter) model presented in section 2 of the main text and simulated two separate populations with different $N$ and $\bar{S}$. The combination of the two simulated status distribution provides a good fit to the American household income distribution. This is shown in Fig \ref{fig:USA_fit_2populations}, where the blue curve is the American household income distribution data for the year 2015, the red and green curves represent the two independently-simulated status distributions, and the cyan curve shows the combination of the two simulated distributions. The simulation that produced the green curve contained $N_1=0.1N_{dat}$ individuals, where $N_{dat}=$ 1,226,728 is the number of households in the dataset. For this first simulated society, $\bar{S}_1=225,000$. The simulation that produced the red curve contained $N_2=0.9N_{dat}$ individuals with $\bar{S}_2=64,445$. $\bar{S}_1$ and $\bar{S}_2$ were chosen so that the total status of the two simulated systems $\bar{S}_1 N_1 + \bar{S}_2 N_2$ was equal to the total household income reported in the dataset.

\paragraph{} As shown in Fig \ref{fig:USA_fit_2populations}, the American household income data is well-represented by the combination of the status distributions of (i)  a simulated system with a relatively small number of individuals having a relatively high average status, and (ii) a simulated system with a relatively large number of individuals having a relatively modest average status. This approach differs from those presented in sections 2.1 of the main text and section \ref{section:SI_3condition_model}, by assuming that the society consists of two separate groups for which the members of each group engage in status-determining interactions amongst themselves, but for which there are no cross-group interactions. The extended model presented in section 2.1 of the main text makes no such assumptions.

\clearpage


\end{document}